\documentclass{revtex4-2}
\usepackage{hyperref}
\usepackage{amsthm,amsmath,amssymb}
\usepackage{setspace}
\usepackage{color}
\usepackage{polyglossia}
\setmainlanguage{english}
\usepackage{physics,latexsym}
\usepackage{graphicx}
\usepackage{tabularx}
\usepackage[labelfont={sf}]{caption}
\usepackage{slashed}
\usepackage[a4paper, total={6.5in, 10in}]{geometry}
\usepackage{makecell}
\usepackage{subcaption} 
\begin{document}

\title{Thermodynamics and Stability of Ultraspinning Black Holes}

\begin{abstract}
    Ultraspinning black holes have attracted considerable attention due to their super-entropic nature, and previous analyses—mostly restricted to neutral cases and high-temperature regimes—have suggested that such black holes are always thermodynamically unstable.
In this work, we revisit the thermodynamic stability of ultraspinning black holes by performing a systematic analysis of the heat capacity in different ensembles over the full range of the horizon radius $r_H$, 
which were missed in earlier temperature-based analyses.
We demonstrate for the first time that, contrary to earlier claims, ultraspinning black holes can admit thermodynamically stable regions, whose existence crucially depends on the spacetime dimension, the solution branch, and the presence of charge.
In addition, we present the first application of the revised reverse isoperimetric inequality to ultraspinning black holes.
Despite the violation of the original reverse isoperimetric inequality 
in this super-entropic regime, 
we find that the revised inequality remains applicable 
and imposes nontrivial constraints on the allowed parameter space, 
including an upper bound on the ultraspinning parameter $\mu$, 
strengthened lower bounds on the mass $m$, 
and upper bounds on both the charge $q$ and the AdS radius $l$. 
To ensure the consistency of the thermodynamic description, 
the conserved charges and the first law in the ultraspinning limit are derived using the Iyer–Wald formalism together with integrability conditions.

\end{abstract}

\author{Zhenbo Di}
\thanks{Corresponding author}
\email{dizhenbo@mail.bnu.edu.cn}
\affiliation{School of Physics and Astronomy, Beijing Normal University, Beijing ,
China,100875}


\maketitle

\tableofcontents

\section{Introduction}
    
Black hole thermodynamics provides an important framework to study the quantum properties of black holes through macroscopic quantities
\cite{PhysRev.164.1776.Israel.1967,bardeen_four_1973.Carter.Hawking,P.C.W.Davies.1978,Carlip.review.2014}.
Beyond conceptual importance, black hole thermodynamics also provides heat capacity as a tool to assess thermodynamic stability of black holes\cite{Monteiro.2010.Stability.nodate,Avramov.review.2024}.
. 
For example, the negative heat capacity of Schwarzschild black holes means that after radiation they will get hotter, which will make the radiation faster.
Thus instability shows up. 

In recent years, black hole thermodynamics with a variable cosmological constant, 
often referred to as thermodynamics in extended phase space or black hole chemistry, has attracted increasing attention
\cite{Kastor2009,Dolan.cosmological.2011,Dolan.pressure.2011,Kubiznak.Mann.2012,Mann.review.2025}.
Treating the cosmological constant as a thermodynamic pressure leads to a variety of novel phenomena, such as phase transitions and rich equation-of-state behavior
\cite{Kubizňák.Mann.2015,Mann.2016.book,Caceres.2017,mann.black.2015.book,gibbons.first.2005,karch.holographic.2015}.
Within this framework, thermodynamic stability is still closely connected to heat capacity, whose sign and divergent structure encodes information of 
locally stable and unstable phases.

Ultraspinning black holes \cite{hennigar_entropy.Hennigar.2015} form a special class of solutions characterized by non-compact horizons and the violation of the reverse isoperimetric inequality (RII).
The latter is the reason why ultraspinning belongs to a kind of so-called 'super-entropic' black holes.
Motivated by these unusual properties, extensive studies have been carried out, 
including the application of the ultraspinning limit to various black hole solutions
\cite{WuDi:2021vch,wu_aspects_Wu.2021}.
In Ref.~\cite{Johnson.2020}, a conjecture was proposed, claiming that all super-entropic black holes are thermodynamically unstable.
Subsequently, counterexamples to this instability conjecture have been identified in lower-dimensional settings,
most notably in exotic BTZ black holes \cite{Cong.Mann.2019},
indicating that super-entropicity does not necessarily imply thermodynamic instability.
Nevertheless, for ultraspinning Kerr-AdS black holes in higher dimensions,
the thermodynamic implications of the ultraspinning limit remain subtle,
and a clear understanding of their stability properties is still to be achieved.

It is worth noting that most of the existing investigations focus on relatively restricted regimes,
typically the high-temperature and neutral cases, where the thermodynamic behavior appears simpler
and the stability properties can be characterized in a relatively straightforward manner.
While such simplifications are technically convenient, they may overlook important structures
in the full parameter space of ultraspinning black holes.

In fact, existing analyses have not provided a complete characterization of the heat capacity
of ultraspinning black holes.
In particular, stability was often evaluated solely as a function of temperature,
which ignores the degeneracy between black hole parameters and thermodynamic quantities.
Moreover, a systematic investigation of the heat capacity for both neutral and charged
ultraspinning black holes is still lacking.

In this work, we present a systematic analysis of the thermodynamic stability
of ultraspinning black holes by explicitly examining their heat capacities.
Contrary to earlier expectations, we find that ultraspinning black holes
admit thermodynamically stable regions in the parameter space. 
The origin of these stable regions can be traced back to a more complete
treatment of the parameter space.
In contrast to earlier analyses, which characterized stability primarily
as a function of temperature and were therefore insensitive to the underlying
branch structure, we adopt the horizon radius $r_H$ as the fundamental variable,
allowing the full branching behavior of thermodynamic quantities to be resolved.
Moreover, previous investigations were largely restricted to neutral black holes
in the high-temperature regime.
Here we extend the analysis to the full allowed range of horizon radii
and include both neutral and charged ultraspinning black holes.

As a consistency check, we recover that the isobaric heat capacity of neutral
ultraspinning black holes is negative for sufficiently large horizon radii
(or equivalently, at sufficiently high temperatures), in agreement with earlier results.
Nevertheless, our analysis reveals the emergence of stable regions in other parts
of the parameter space, both in different spacetime dimensions and in the presence
of electric charge.

As a complementary analysis, we also examine the revised reverse isoperimetric inequality (RRII) , which was recently 
proposed in the literature \cite{PhysRevLett.131.241401.Amo.2023}. 
The RRII can be applied to ultraspinning cases and it becomes constraints between range of black hole parameters. 
In particular, 
the RRII yields additional and independent constraints on the parameter space of ultraspinning black holes. 
Our analysis indicates that it provides an upper bound on the ultraspinning parameter $mu$, 
suggests a strengthened lower bound on the mass parameter $m$ compared with that required by the existence of a horizon, 
and gives upper bounds on both the charge $q$ and the AdS radius $l$. 
Since the prove of RRII is still lacking, here we treat it as an assumption. 
Thus the content of RRII is independent with the analyses on heat capacities and local stability.

To evaluate the thermodynamic quantities, we adopt the Iyer–Wald formalism \cite{PhysRevD.48.R3427.Wald93,PhysRevD.52.4430.Wald.1995,PhysRevD.50.846.Wald94} 
together with an integrability prescription developed in Ref.~\cite{Gao:2023luj.Yunjiao.Gao.2023}, 
which provides a consistent definition of conserved charges in asymptotically AdS spacetime.
Our results agree with those obtained using the conformal completion method\cite{hennigar_entropy.Hennigar.2015}, offering a nontrivial consistency check of the formalism in the ultraspinning regime.


The paper is organized as follows.
In Sec.~II, we briefly review ultraspinning black hole solutions and outline the Iyer–Wald formalism with the integrability prescription, 
then these methods are applied to ultraspinning black holes to evaluate their thermodynamics.  
In Sec.~III, we analyze the heat capacity in detail and investigate the thermodynamic stability.
Sec.~IV contains brief introductions to entropy inequalities and the results of constraining parameters from RRII. 
Sec.~V contains our conclusions and further discussion. 
Technical details and complementary discussions are collected in the appendices.

    \section{Ultraspinning Black Holes and Its Thermodynamics from Iyer-Wald Formalism}
    In this section, 
    we briefly review the construction of ultraspinning black holes 
    and summarize the essential ingredients of the Iyer–Wald formalism 
    together with the integrability prescription adopted in the extended phase space. 
    These tools are then applied to ultraspinning black holes to obtain the corresponding thermodynamic quantities. 
    Finally, the results are independently checked using the conformal completion method.

    \subsection{Ultraspinning Black Holes}
    Ultraspinning black holes can be obtained from Kerr-Newman-AdS black holes by taking the limit of $a \rightarrow l$.
    This kind of black holes has received attention because of its non-compact horizon \cite{klemm_four-dimensional_Klemm.2014} 
    and its violation to reverse isoperimetric inequality (RII) \cite{hennigar_entropy.Hennigar.2015}, which was checked to be true for a large amount of black holes \cite{Pope.Gibbons.RII.PRD}. 
    In \cite{PhysRevLett.131.241401.Amo.2023}, the authors propose a revised version of RII, it worthy checking the revised version for ultraspinning case.
    The violation and constraints from RRII can be found in Appendix.

    Metric of Kerr-Newman-AdS black hole can be written in standard Boyer-Lindquist coordinate\cite{Hawking_rotation_1998}:
    \begin{equation}
        \begin{aligned}
            ds^{2} &=-\frac{\Delta_{a}}{\Sigma_{a}}\left[dt-\frac{a\mathrm{sin}^{2}\theta}{\Xi}d\phi\right]^{2}+\frac{\Sigma_{a}}{\Delta_{a}}dr^{2}+\frac{\Sigma_{a}}{S}d\theta^{2} 
            +\frac{S\mathrm{sin}^{2}\theta}{\Sigma_{a}}\left[adt-\frac{r^{2}+a^{2}}{\Xi}d\phi\right]^{2}, \\
            &\mathcal{A}=-\frac{qr}{\Sigma_{a}}\left(dt-\frac{a\mathrm{sin}^{2}\theta}{\Xi}d\phi\right),
        \end{aligned}
    \end{equation}

    where
    \begin{equation}
        \begin{aligned}
            &&&&&\Sigma_{a} =r^2+a^2\mathrm{cos}^2\theta,\quad\Xi=1-\frac{a^2}{l^2},  \\
            &&&&&S=1-\frac{a^{2}}{l^{2}}\mathrm{cos}^{2}\theta,  \\
            &&&&&\Delta_{a} =(r^2+a^2)\left(1+\frac{r^2}{l^2}\right)-2mr+q^2 ,
        \end{aligned}
    \end{equation}
    in which $m$ and $a$ are parameters related to mass and angular momentum separately, $l$ is AdS radius.
    To get the ultraspinning black holes, coordinate $\phi$ should be changed to $\phi = \frac{\psi}{\Xi} $, then take the limit: $a \rightarrow l$,
    in this way, one get the ultraspinning solution, which is also a solution of Einstein-Maxwell equation.
    \begin{equation}
        \begin{aligned}
            ds^{2}& =-\frac\Delta\Sigma\left[dt-l\mathrm{sin}^2\theta d\psi\right]^2+\frac\Sigma\Delta dr^2+\frac\Sigma{\mathrm{sin}^2\theta}d\theta^2 
            +\frac{\sin^{4}\theta}\Sigma[ldt-(r^{2}+l^{2})d\psi]^{2}, \\
            A& =-\frac{qr}\Sigma(dt-l\mathrm{sin}^2\theta d\psi), 
        \end{aligned}
    \end{equation}
    where
    \begin{equation}
        \Sigma=r^2+l^2\mathrm{cos}^2\theta,\quad\Delta=\left(l+\frac{r^2}{l}\right)^2-2mr+q^2.
    \end{equation}
    The range of $\psi$ becomes non-compact when $a \rightarrow l$ ( which comes from $\Xi = 1-\frac{a^2}{l^2} \rightarrow 0$).
    One can compactify $\psi$ coordinate as $\psi \sim \psi + \mu$.

    The horizon locates at $r_h$, which is the largest root of $\Delta(r) = 0$. 
    And the horizon is topologically non-compact, can be viewed as a sphere with two punctures.

    And the parameters must obey below condition to make horizon exists:
        \begin{equation}
            m \geq 2 r_0 \left(\frac{r_0^2}{l^2} + 1\right) ;\quad r_0^2 = \frac{l^2}{3}\left[-1 + \left(4 + \frac{3}{l^2} q^2\right)^{1/2}\right].
        \end{equation}

    \subsection{A Brief Review of Iyer-Wald Formalism}
    The Iyer-Wald formalism is one of the standard analytical methods in black hole thermodynamics, 
    in which thermodynamic quantities are evaluated and interpreted in a geometric manner. For example, 
    entropy is identified with the Noether charge associated with Killing vector which is tangent to the horizon.

    For black holes with cosmological constant, this method still works, but requires certain modification \cite{Gao:2023luj.Yunjiao.Gao.2023}. 
    Here we present a brief review of this approach and more details are deferred to appendices.

    The Einstein-Hillbert-Maxwell Lagrangian with a cosmological constant can be written as:
        \begin{equation}
            \boldsymbol{L} = \frac{1}{16 \pi}(R - 2 \Lambda - F^{ab}F_{ab})\boldsymbol{\epsilon},
            \label{LEHM}
        \end{equation}
    where the bold symbols denote differential forms, $R$ is the Ricci scalar, $\boldsymbol{F} = d \boldsymbol{A}$ is the electromagnetic field strength tensor,
    $\boldsymbol{\epsilon}$ is the associated volume form, which satisfies $\nabla_a \boldsymbol{\epsilon}=0$, 
    $\Lambda$ is the cosmological constant.
       
    Under a variation of $\boldsymbol{L}$, one finds:
    $$\delta \boldsymbol{L} = \boldsymbol{E}_{g} \delta g^{ab} + \boldsymbol{E}_{A} \delta A_a - \frac{1}{8\pi} \boldsymbol{\epsilon}\delta \Lambda + d\boldsymbol{\Theta},$$
    
    where all $\boldsymbol{E}$ represent tensors associated to equations of motion, 
    $d\boldsymbol{\Theta}$ is the so-called boundary term, which will be used to construct Noether charge, and $\Lambda$ is also viewed as a variable.

    After evaluating boundary term from the Lagrangian, one can construct Noether current from it:
    \begin{equation}
        \boldsymbol{J} = \boldsymbol{\Theta}(\Psi,\mathcal{L}_{\xi}\Psi) - \xi \cdot \boldsymbol{L} = \boldsymbol{C}_{\xi} - d \boldsymbol{Q}
        \label{J}
    \end{equation}
    Here $\xi^a$ is the vector field associated with diffeomorphism invariance, $\Psi$ represents dynamical fields such as $g_{ab}$ and $A_a$.

    We define a new variation operator called $\bar{\delta}$, which means not varying $\xi^a$ and the relation of $\delta$ and $\bar{\delta}$ is $\bar{\delta} \boldsymbol{X}_{\xi} = \delta \boldsymbol{X}_{\xi} - \boldsymbol{X}_{\delta \xi}$.
    Then the variation of $\boldsymbol{J}$ is:
    \begin{equation}
        \begin{aligned}
            \bar{\delta} \boldsymbol{J} & = \bar{\delta} \boldsymbol{\Theta} - \xi \cdot  \bar{\delta} \boldsymbol{L}\\
            & = \boldsymbol{\omega}(\Psi,\bar{\delta}\Psi,\mathcal{L}_{\xi}\Psi) - \xi \cdot \boldsymbol{E}_{\psi}\delta \psi  +  d(\xi \cdot \Theta) + \frac{1}{8 \pi}\xi \cdot \boldsymbol{\epsilon} \delta \Lambda.
        \end{aligned}
    \end{equation}
    When $\xi$ in above expression is a Killing vector field as well as representing a symmetry of dynamical fields 
    and EOMs hold, i.e. $\mathcal{L}_{\xi}\psi = 0$ and $\boldsymbol{C}_\xi =0$, $\omega(\psi,\bar{\delta}\psi,\mathcal{L}_{\xi}\psi) = 0$ \cite{PhysRevD.50.846.Wald94},
    the above formula can be simplified to:
    \begin{equation}
        d(\bar{\delta}\boldsymbol{Q}_\xi - \xi \cdot \boldsymbol{\Theta}) = \frac{1}{8\pi} \xi \cdot \boldsymbol{\epsilon}\delta \Lambda.
    \end{equation}
    
    In the presence of cosmological constant, the method differs slightly from asymptotic flat case. 
    Integrate both sides from horizon (denoted as $H$) to any surface of $t, r=constant$ living out of horizon, denoted as $S$, then apply Stokes' theorem to LHS and separate RHS into two parts:
    \begin{equation}
        \int_S \bar{\delta}\boldsymbol{Q}_\xi - \xi \cdot \boldsymbol{\Theta} - \int_H \bar{\delta}\boldsymbol{Q}_\xi - \xi \cdot \boldsymbol{\Theta} = \int_{r=0}^{r=r_s} \frac{1}{8\pi} \xi \cdot \boldsymbol{\epsilon}\delta \Lambda - \int_{r=0}^{r=r_H} \frac{1}{8\pi} \xi \cdot \boldsymbol{\epsilon}\delta \Lambda.
    \end{equation}

    From above relation, one readily finds that there exists a quantity is independent of the choice of integration surface:
    \begin{equation}
        \bar{\delta} \int_S \boldsymbol{Q}_\xi - \int_S\xi \cdot \boldsymbol{\Theta} - \int_{r=0}^{r=r_s} \frac{1}{8\pi} \xi \cdot \boldsymbol{\epsilon} \delta \Lambda.
    \end{equation} 
    When the Killing vector is chosen to be $t^a \equiv \left(\pdv{}{t}\right)^a$ and $-\phi^a \equiv -\left(\pdv{}{\phi}\right)^a$, 
    the associated conserved charges are denoted by $\slashed{\delta}\mathcal{M}$ and $\delta J$.
    Fix the gauge conditions $\delta t^a = 0$ and $ \delta \phi^a = 0$, and 
    consider Killing vector $\xi^a = t^a + \Omega_H \phi^a$ ($\Omega:= - \frac{g_{t \phi}}{g_{\phi \phi}} |_H$, angular velocity of black hole).

    With appropriate transformations and definitions (more details can be found in Appendix), 
    the first law with $P-V$ term can be deduced:
    \begin{equation}
        \slashed{\delta} \mathcal{M} = \frac{\kappa}{8\pi}\delta A +\Omega_H \delta J +\Phi_H \delta Q + V \delta P.
        \label{FirstLaw}
    \end{equation}

    \subsection{Thermodynamics of Ultraspinning Black Holes}
    In this subsection, 
    the thermodynamics of ultraspinning black holes is revisited 
    by evaluating the relevant thermodynamic quantities, 
    the first law, 
    and the Smarr relation 
    using the Iyer–Wald formalism 
    together with the integrability method. 
    This analysis demonstrates the effectiveness of this approach in the ultraspinning regime.

    \subsubsection{Thermodynamic Quantities from Definitions}
    The thermodynamic quantities (except for the mass) can be evaluated directly from their definitions \ref{def thermo quantities_A}: 
\begin{equation}
    \begin{aligned}
        A & := \int_{S} \sqrt{\sigma}, & \kappa & := \sqrt{-\frac{1}{2} \nabla^{a} \xi^b \nabla_{a} \xi_b }, \\
        J &:= -\int_S \boldsymbol{Q}_\psi, & \Omega &:= - \frac{g_{t \psi}}{g_{\psi \psi}} |_H; \\
        Q &:= \frac{1}{8\pi} \int_H F^{ab} \epsilon_{abcd}, & \Phi_H &:= -A_a \xi^a|_H , \\
        V &:= \int_{r=0}^{r=r_H}t \cdot \boldsymbol{\epsilon}, &  P & := -\frac{ \Lambda}{8 \pi} ,
    \end{aligned}
    \label{def thermo quantities}
\end{equation}
Apply these definitions to ultraspinning black hole solution, it is directly to obtain:
\begin{equation}
    \begin{aligned}
        A & = 2 \mu (l^2 + r_H^2),   & &\kappa = \frac{1}{2 r_H}(3\frac{r_H^2}{l^2} -1 -\frac{q^2}{l^2 + r^2_H}),\\
        J & = \frac{l m \mu}{2 \pi}, & &\Omega_H = \frac{l}{r_H^2 + l^2},\\
        Q & = \frac{q \mu}{2 \pi},   & &\Phi = \frac{q r_H}{l^2 + r_H^2},\\
        V & = \frac{2}{3} \mu r_H (r_H^2 + l^2), &  &P = \frac{3}{8 \pi l^2},
    \end{aligned}
\end{equation}

\subsubsection{Integrable Mass}
    After evaluating the thermodynamics of black hole, the freedom in choosing the Killing vector can be exploited to make the first law integrable, 
    as was done for the Kerr-AdS case in  \cite{Gao:2023luj.Yunjiao.Gao.2023}. 

    The first law obtained from Iyer-Wald Formalism reads (Since the first law is not integrable, the variation of mass is denoted as $\slashed{\delta} \mathcal{M}$): 
\begin{equation}
\begin{aligned}
    \slashed{\delta} \mathcal{M} & = \frac{\kappa}{8 \pi} \delta A + \Omega_H \delta J + \Phi_H \delta Q + V \delta P \\
    & = \mu \frac{\delta m}{2 \pi} + \frac{m (l^2 + 3 r^2_H)}{4\pi (l^2 + r^2_H)} \delta \mu. 
\end{aligned}
\end{equation}

The above result cannot be made integrable by multiply an overall factor. 
Because the coefficient of $\delta \mu$ is a function of $l$,
but after evaluation, the coefficient of $\delta l$ is found to be 0. 
Regardless of the factor introduced, the missing $\delta l$ term cannot be added back. 

In order to restore integrability of the first law, a subtraction of $K \delta \mu$ term is introduced:
\begin{equation}
\slashed{\delta} \mathcal{M} = \delta \left(\frac{\mu m}{2\pi}\right) - K \delta \mu,
\end{equation}
\begin{equation}
    K = \frac{m (l^2 - r_H^2)}{4 \pi (l^2 + r_H^2)} = \frac{(l^2 - r^2_H) \left[(l^2 + r^2_H)^2+q^2 l^2\right]}{8 \pi l^2 r_H (l^2 + r_H^2)}.
\end{equation}

And the integrable first law is: 
\begin{equation}
\delta \mathcal{\tilde{M}} = \frac{\kappa}{8\pi}\delta A +\Omega_H \delta J +\Phi_H \delta Q + V \delta P + K\delta \mu,
\end{equation}
in which $\delta \mathcal{\tilde{M}}$ represents an integrable mass:
\begin{equation}
\mathcal{M} = \frac{\mu m}{2 \pi}.
\end{equation}
The $K \delta \mu$ term admits a natural interpretation as a chemical potential contribution \cite{PhysRevLett.115.031101.Mann.Hennigar.2015}. 
Since coordinate $\psi$ becomes null when $r \rightarrow \infty$, 
which makes $\mu$ become period of a null coordinate.
According to the discussion in \cite{herzog_heating_2008}, 
the conserved charge associated with a null vector other than time translation is number of particles. 
In the field theory living on boundary, 
the conserved charge associated with $\pdv{}{\psi}$ can be explained as number of particles, 
and $\mu$, which is conjugate to the conserved charge, can be explained as chemical potential.

The results from Iyer-Wald formalism and integrable method are consistent with the result in \cite{PhysRevLett.115.031101.Hennigar.Mann2016} by conformal completion method.

\subsubsection{Higher Dimensional Case}
For higher dimensional ultraspinning black holes, a similar procedure can be applied. 

Higher dimensional single angular momentum ultraspinning black holes can be obtained from higher dimensional Kerr - AdS black holes by
similar method.

The metric of higher dimensional single angular momentum Kerr - AdS black holes is\cite{Hawking_rotation_1998}:
\begin{equation}
\begin{aligned}
    ds^{2}&=-\frac{\Delta_{r}}{\rho^{2}}\left(dt-\frac{a}{\Xi}\operatorname{sin}^{2}\theta d\phi\right)^{2}+\frac{\rho^{2}}{\Delta_{r}}dr^{2}+\frac{\rho^{2}}{\Delta_{\theta}}d\theta^{2}\\
   & \qquad +\frac{\Delta_\theta\sin^2\theta}{\rho^2}\left[adt-\frac{(r^2+a^2)}{\Xi}d\phi\right]^2+r^2\cos^2\theta d\Omega_{d-4}^2, \\
   \Delta & _r=(r^2+a^2)(1+l^2r^2)-2 M r^{5-d}, \\
   \Delta_{\theta}& =1-\frac{a^2}{l^2}\cos^2\theta, \\
   \Xi & =1-\frac{a^2}{l^2}; \\
   \rho^{2}& =r^2+a^2\cos^2\theta. 
\end{aligned}
\end{equation}
Redefine the $\phi$ coordinate: $\psi = \frac{\phi}{\Xi}$, under the limit: $a \rightarrow l$, and use $\psi \sim \psi + \mu$ to compactify this metric: 
\begin{equation}
\begin{aligned}
    ds^{2}& =-\frac\Delta{\rho^2}(dt-l\mathrm{sin}^2\theta d\psi)^2+\frac{\rho^2}\Delta dr^2+\frac{\rho^2}{\mathrm{sin}^2\theta}d\theta^2 \\
    &+\frac{\sin^4\theta}{\rho^2}[ldt-(r^2+l^2)d\psi]^2+r^2\mathrm{cos}^2\theta d\Omega_{d-4}^2, \\
\end{aligned}
\end{equation}
\begin{equation}
\Delta=\left(l+\frac{r^2}l\right)^2-2 m r^{5-d},\quad\rho^2=r^2+l^2\mathrm{cos}^2\theta.
\end{equation}
The term $d \Omega_{d-4} ^2$ represents the sphere unit of $d-4$dimension sphere, denote the angular coordinates as $\{\alpha_i\}$ and $i = 1,2,\cdots,d-4$, the metric of this sphere can be written as:
\begin{equation}
\text{d}\Omega_{d-4}^2 = \text{d} \alpha_1 ^2 + \sin^2 \alpha_1 \text{d} \alpha_2^2 + \cdots + \sin^2 \alpha_1 \sin^2 \alpha_2 \cdots \sin^2 \alpha_{d-5} \text{d}\alpha_{d-4}^2 ,
\end{equation}
with each angle taking values in $\left[\right.0,\pi\left.\right)$

By definition \ref{def thermo quantities}, 
the thermodynamic quantities can be evaluated: 
\begin{equation}
\begin{aligned}
    & T = \frac{(d-1) r_H^2+(d-5) l^2}{4 \pi  l^2 r_H} ;& & S =  \frac{A}{4} = \frac{\pi ^{\frac{d-3}{2}} \mu  r_H^{d-4} \left(r_H^2+l^2\right)}{4 \Gamma \left(\frac{d-1}{2}\right)}\\
    & V = \frac{\pi ^{\frac{d-3}{2}} \mu  r_H^{d-3} \left(r_H^2+l^2\right)}{2 \Gamma \left(\frac{d+1}{2}\right)};& & P = \frac{(d-2) (d-1)}{16 \pi  l^2}\\
    & \Omega_H = \frac{l}{r_H^2+l^2} ;& & J = \frac{\pi ^{\frac{d-5}{2}} l \mu M}{2 (d-3) \Gamma \left(\frac{d-3}{2}\right)}
\end{aligned}
\end{equation}

Only when the first law is integrable——so that the variation $\delta$ can be treated as a exact differential symbol,
one can apply scaling argument to get Smarr formulation. And the freedom of choosing Killing vector makes this practical.

After that, integrability procudere should be applied to extract $\delta \mu$ term and get the integrable mass.

The variation of mass from Iyer-Wald Formalism is: 
\begin{equation}
\slashed{\delta} \mathcal{M} = \frac{\pi ^{\frac{d-5}{2}}   r_H^{d-5} \left(r_H^2+l^2\right) \left((d-1) r_H^2+(d-3) l^2\right)}{16 l^2 \Gamma \left(\frac{d-1}{2}\right)}\delta \mu + \frac{(d-2) \pi ^{\frac{d-5}{2}} \delta m \mu }{8 \Gamma \left(\frac{d-1}{2}\right)}.
\end{equation}
By extracting a $\delta \mu$ term, 
\begin{equation}
\mathcal{M} = \frac{(d-2) \pi ^{\frac{d-5}{2}} \mu  m}{8 \Gamma \left(\frac{d-1}{2}\right)}.
\end{equation}
and the conjugate quantity of $\mu$ is: 
\begin{equation}
K = -\frac{\delta \mathcal{M} - \delta \tilde{\mathcal{M}}}{\delta \mu} = \frac{\pi ^{\frac{d-5}{2}} r_H^{d-5} \left(l^4-r_H^4\right)}{16 l^2 \Gamma \left(\frac{d-1}{2}\right)}.
\end{equation}

\subsubsection{Constraints on the Choice of Killing Vector from Integrability}
    The requirement of integrable first law provides an additional contraint, 
    which can be used to constrain Killing vecter associated with charges.

    In many approaches to define the mass of black hole, one need to choose a timelike Killing vector, and evaluate a conserved charge associated to it.

    For asymptotically flat case, a natural constrain can be applied to the Killing vector, which require it to be unit.

    But for asymptotically AdS spacetime, this convenient requirement cannot be hold. Thus the choice of Killing vector in asymptotically AdS can only be determined 
    up to a coefficient. Many earily works simply choose $\pdv{}{t}$, following the habit from asymptotically flat case. 
    But in \cite{Gao:2023luj.Yunjiao.Gao.2023}, the authors' proposal is, for Kerr - AdS black holes, one can use integrable method to further constrain the Killing vector.

    Here, we apply the same method to check if this method can give constraint for Killing vector in ultraspinning black holes cases.
\begin{equation}
    \begin{aligned}
        \slashed{\delta} \mathcal{M}'(\beta) & = \bar{\delta} \int_{H} \boldsymbol{Q}[\beta t] - \int_{H} \beta t \cdot \boldsymbol{\Theta} - \int_{r = 0}^{r = r_H} \frac{1}{8 \pi} \beta t \cdot \boldsymbol{\epsilon} \delta \Lambda \\ 
                 & = \beta \left(\mu \frac{\delta m}{2 \pi} + \frac{  m \left(l^2+3 r_H^2\right)}{4 \pi  \left(l^2+r_H^2\right)}\delta \mu\right).
    \end{aligned}
\end{equation}

The next step is to constrain $\beta(l,q,\mu)$ using integrability. 
To cover the most general situation, $\beta$ is assumed to be a function of $q,l,\mu$.

Similar evaluation method is used to get chemical potential for $\beta \pdv{}{t}$ case:
\begin{equation}
    \slashed{\delta} \mathcal{M}' = \beta \slashed{\delta} \mathcal{M} = \delta{\mathcal{M}'} - K' \delta \mu 
    = \pdv{\mathcal{M}'}{M} \delta M + \pdv{\mathcal{M}'}{l} \delta l + \pdv{\mathcal{M}'}{q} \delta q + (\pdv{\mathcal{M}'}{\mu} - K') \delta \mu.
\end{equation}
Compare this result and the result from direct variation of $\tilde{M'}$ and $K'$:
\begin{equation}
    \begin{aligned}
        & \pdv{\mathcal{M}'}{M} = \frac{\beta \mu}{2 \pi} , \\
        & \pdv{\mathcal{M}'}{\mu} - K' = \beta \frac{M (l^2 + 3 r_H^2)}{4 \pi (l^2 + r_H^2)} , \\
        & \pdv{\mathcal{M}'}{l} = 0 = \pdv{\tilde{\mathcal{M}}'}{q}. 
    \end{aligned}
\end{equation}
It is direct to get:
\begin{equation}
    \begin{aligned}
        \mathcal{M}' & = \frac{\beta \mu m}{2 \pi}, \\
        K' & =  \frac{\left(l^2 \left(2 r_H^2+q^2\right)+r_H^4+l^4\right) \left(2 \mu  \beta '(\mu ) \left(r_H^2+l^2\right)+\beta (\mu ) \left(l-r_H\right) \left(r_H+l\right)\right)}{8 \pi  l^2 r_H \left(r_H^2+l^2\right)}.
    \end{aligned}
\end{equation}
the coefficients of $\delta l$ and $\delta q$ vanish, which indicates that $\beta$ should be independent of $q$ and $l$.
Thus $\beta$ should only be a function of $\mu$.

\subsubsection{Scaling Argument and Smarr Relation}
Scaling argument can be applied to get the relationship between thermodynamic quantities, which is known as Smarr relation.

A scaling transformation can be applied to metric :
$g_{ab} \rightarrow \alpha^2 g_{ab}$. 
After that, the thermodynamic quantities transform as:
\begin{equation}
\begin{aligned}
    &\mathcal{M} \rightarrow \alpha^{d-3} \mathcal{M} ,\\
    &\kappa \rightarrow \alpha^{-1} \kappa , A \rightarrow \alpha^{d-3} A, \\
    &\Omega_H \rightarrow \alpha^{-1} \Omega_H , J \rightarrow \alpha^{d-2} J, \\
    &V \rightarrow \alpha^{d-1} V , P \rightarrow \alpha^{-2} P , \\
    &K \rightarrow \alpha^{d-3} K , \mu \rightarrow \mu.
\end{aligned}
\end{equation} 

Bring the results back to the integrable first law,

\begin{equation}
\delta (\alpha^{d-3} \mathcal{M}) = \frac{\alpha^{-1} \kappa }{2\pi} \delta(\frac{\alpha^{d-3} A}{4}) + \alpha^{-1} \Omega_H \delta (\alpha^{d-2} J)
    + \alpha^{d-1} V \delta (\alpha^{-2} P ) + \alpha^{d-3} K \delta \mu.
\end{equation}
Since a scaling transformation does not alter the physical content, first law should keep unchanged before and after the transformation, 
that is why the coefficient of $\delta \alpha$ term should be 0, 
which gives:
\begin{equation}
\frac{d-3}{d-2}\mathcal{M} = -\frac{2}{d-2} P V + T S + \Omega_H J.
\end{equation}

    \subsection{Comparison Between Results from Iyer-Wald Formalism and Conformal Completion Method}
    For $d \geq 4$, we present a comparison of the mass and angular momentum obtained from the Iyer--Wald formalism and the conformal completion method. 
Here $\mathcal{M}_C$ and $J_C$ denote the mass and angular momentum evaluated using the conformal completion approach, respectively.

As shown in Table~\ref{tab:comparison}, the conserved charges obtained from the Iyer--Wald formalism are in exact agreement with those derived using the conformal completion method for all dimensions considered.

\begin{table}[htbp]
\centering
\renewcommand{\arraystretch}{1.4}
\begin{tabular}{|c|c|c|c|c|c|c|}
\hline
$d$ & $\mathcal{E}_t^{\; t}$ & $\mathcal{E}_\psi^{\; t}$           
& $\mathcal{M}_C$                           
& $ J_C$                        
& $\tilde{\mathcal{M}}$   
& $\tilde{J}$ \\
\hline
4   
& $2 m$            
& $-3 l m \sin ^2(\theta )$ 
& $\frac{\mu m}{2\pi}$       
& $\frac{ \mu m l}{2 \pi}$       
& $\frac{\mu m}{2\pi}$       
& $\frac{ \mu m l}{2 \pi}$  \\
\hline
5   
& $6 m$            
& $-8 l m \sin ^2(\theta )$ 
& $\frac{3 \mu m}{8}$       
& $\frac{ \mu m l}{4}$       
& $\frac{3 \mu m}{8}$       
& $\frac{ \mu m l}{4}$   \\
\hline
6   
& $12 m$            
& $-15 l m \sin ^2(\theta )$ 
& $\frac{2 \mu m}{3}$       
& $\frac{1}{3}  \mu l m$       
& $\frac{2 \mu m}{3}$       
& $\frac{1}{3}  \mu l m$\\
\hline
7   
& $20 m$            
& $-24 l m \sin ^2(\theta )$ 
& $\frac{5 \pi \mu m}{16}$       
& $\frac{1}{8} \pi \mu l m$       
& $\frac{5 \pi \mu m}{16}$       
& $\frac{1}{8} \pi \mu l m$ \\
\hline
8   
& $30 m$            
& $-35 l m \sin ^2(\theta )$ 
& $\frac{2 \pi \mu m}{5}$       
& $\frac{2}{15} \pi \mu l m$     
& $\frac{2 \pi \mu m}{5}$       
& $\frac{2}{15} \pi \mu l m$  \\
\hline
9   
& $42 m$             
& $-48 l m \sin ^2(\theta)$ 
& $\frac{7}{48} \pi ^2 \mu m$   
& $\frac{1}{24} \pi ^2 \mu l m$   
& $\frac{7}{48} \pi ^2 \mu m$   
& $\frac{1}{24} \pi ^2 \mu l m$\\
\hline
10  
& $56 m$            
& $-63 l m \sin ^2(\theta)$ 
& $\frac{16}{105} \pi ^2 \mu m$ 
& $\frac{4}{105} \pi ^2 \mu  l m$ 
& $\frac{16}{105} \pi ^2 \mu m$ 
& $\frac{4}{105} \pi ^2 \mu  l m$\\  
\hline 
\end{tabular}
\caption{Comparison of the mass and angular momentum obtained from the Iyer--Wald formalism and the conformal completion method in spacetime dimensions $4 \leq d \leq 10$, showing complete agreement between the two approaches. 
$\mathcal{E}_\mu^{\; \nu}$ are the components of the 'electric part' of Weyl tensor used in the conformal completion method.}
\label{tab:comparison}
\end{table}

\newpage
\section{Heat Capacity and thermodynamic Stability of Ultraspinning Black Holes}
There are several distinct notions of stability for black holes. 
From a dynamical perspective, stability is usually assessed by introducing small perturbations 
around a given background solution and examining whether such perturbations grow or decay with time. 
From a thermodynamic viewpoint, stability is instead characterized by response functions, 
among which the heat capacity plays a central role. 
A negative heat capacity implies that the black hole becomes hotter as it loses energy, 
leading to a runaway process that signals thermodynamic instability. 
More generally, the heat capacity encodes how energy is stored in the system 
and thus provides insight into the effective degrees of freedom excited under thermodynamic fluctuations.

In this section, we investigate the thermodynamic stability of ultraspinning black holes by analyzing their heat capacities. In the extended phase space, different ensembles naturally arise, and in particular we consider both the heat capacity at constant pressure 
$C_P$ and at constant volume $C_V$. The former corresponds to fluctuations at fixed cosmological constant (or AdS radius 
$l$) and is therefore of direct physical relevance in AdS black hole thermodynamics. 
The latter probes energy fluctuations without changing the thermodynamic volume 
and provides a more direct measure of the excited degrees of freedom, 
even though the constant-volume ensemble is less straightforward to realize in gravitational systems.

Previous work proposed a conjecture that all super-entropic black holes are thermodynamically unstable \cite{Johnson.2020}, 
which was explicitly verified for the BTZ black hole and for ultraspinning black holes in certain high-temperature regimes. 
Here, we extend this analysis by considering the ultraspinning black hole in a more complete setting, 
including multiple solution branches as well as charged configurations. 
Our results show that both $C_P$ and $C_V$ 
can exhibit highly nontrivial branch-dependent behavior, with sign changes and stable intervals appearing in different parameter regimes.

Throughout this section, we do not impose the reverse isoperimetric inequality, 
as it remains a conjectural criterion rather than a fundamental requirement of black hole thermodynamics. 
Thermodynamic stability is therefore understood as local stability within a given ensemble and solution branch, 
characterized by the sign of the corresponding heat capacity. The existence of multiple branches demonstrates that 
thermodynamic quantities alone do not uniquely determine stability properties: 
ultraspinning black holes with identical macroscopic parameters may nevertheless display qualitatively different stability behavior 
depending on the branch under consideration. 
This enriches and refines previous conclusions on the instability of super-entropic black holes.

\subsection{Lack of van der Waals Phase Transition}

the relation between pressure $P$ and specific volume can be evaluated from thermodynamic quantities,
and has been studied in \cite{hennigar_entropy.Hennigar.2015}:
\begin{equation}
P = \frac{2 \pi T v + 1}{2 \pi v^2},
\end{equation}
in which, $v = \frac{6 V}{A} = 2 r_H$.

and $P-v$ diagrams in different temperature are shown as below:
\begin{figure}[htbp]
\centering
\includegraphics[width=0.4\textwidth]{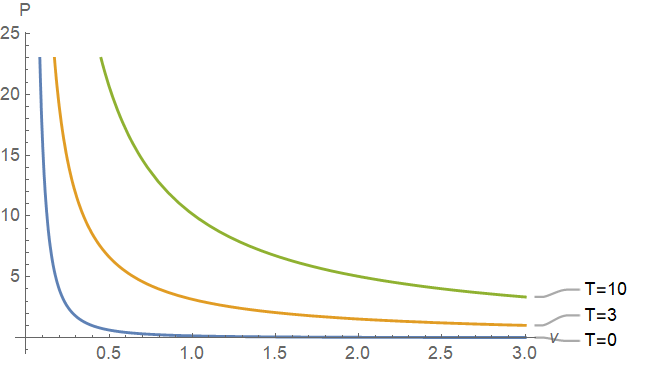}
\caption{$P-v$}
\label{P-v}
\end{figure}

As shown in Fig.\ref{P-v} ultraspinning black holes does not exhibit van der Waals phase transition.

\subsection{Heat Capacity of 4 Dimensional Ultraspinning Black Holes}
In this section, heat capacities of ultraspinning black holes in constant pressure $P$ and constant volume $V$ is evaluated.

To analyze the stability properties unambiguously, $r_H$ is chosen to be variable in order to avoid ambiguity form
temperature as a multivalued function and to observe stability of different size black holes.

\begin{figure}[htbp]
    \centering
    \includegraphics[width=0.4\textwidth]{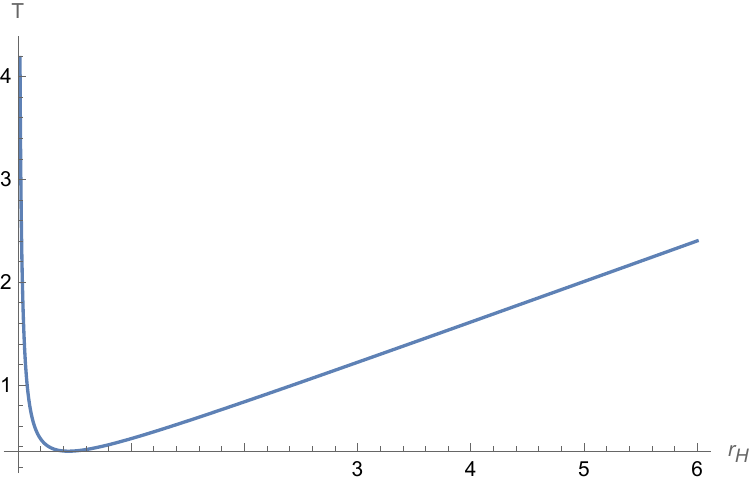}
    \caption{$T-r_H$ diagram.
    \textcolor{black}{From a thermodynamic point of view, the horizon radius provides a monotonic parametrization of the intrinsic geometric scale of the black hole, avoiding ambiguities associated with multi-value temperature branches.}}
\end{figure}

$C_P$ and $C_V$ can be directly evaluated:
\begin{equation}
    \begin{aligned}
        C_P & = T \pdv{S}{T}|_{P,J};\\
        C_V & = T \pdv{S}{T}|_{V,J};\\
    \end{aligned}
\end{equation}

In order to get functions on $r_H$, 
the chain rule is applied and other parameters are replaced to thermodynamic quantities.
Besides,from same thermodynamic quantities, there may exist different solution of black hole parameters 
since $\mu$ as a compactify parameter can be modified. 
Here and below, two distinct solution branches arise. 
We label them as Branch A and Branch B, 
representing solution with larger and smaller $\mu$. 

\begin{figure}[htbp]
\centering
\begin{subfigure}{0.40 \textwidth}
    \centering
    \includegraphics[width=\textwidth]{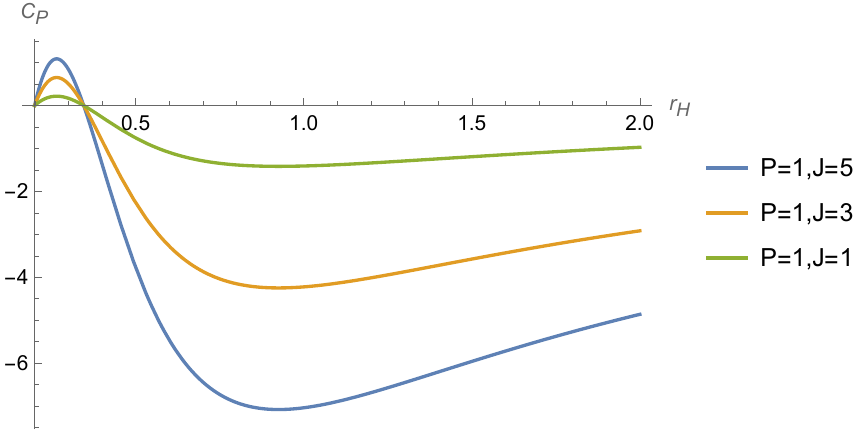}
\end{subfigure}
\begin{subfigure}{0.40 \textwidth}
    \centering
    \includegraphics[width=\textwidth]{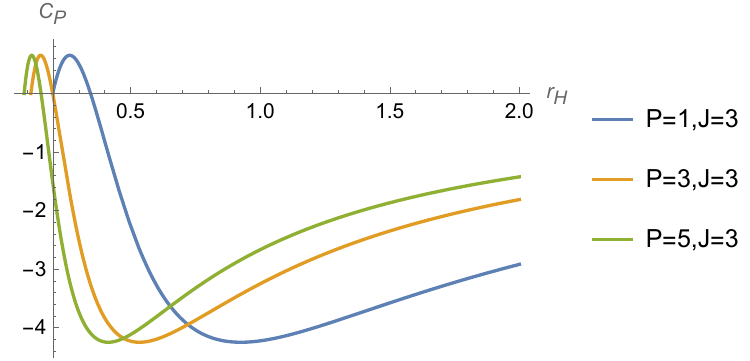}
\end{subfigure}

\caption{$d=4$, $C_P - r_H$ diagram. The range of $r_H$ is chosen to meet $r_H > r_0$ to make sure horizon exists.
}
\label{CP-rH-4}
\end{figure}

\begin{figure}[htbp]
\centering
\centering
\begin{subfigure}{0.4\textwidth}
    \includegraphics[width=\textwidth]{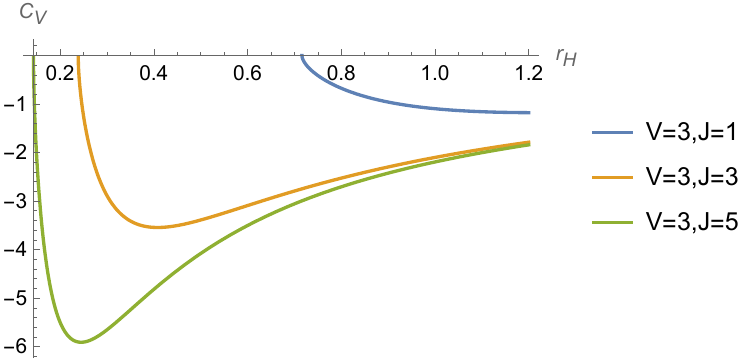}
\end{subfigure}
\begin{subfigure}{0.4\textwidth}
    \includegraphics[width=\textwidth]{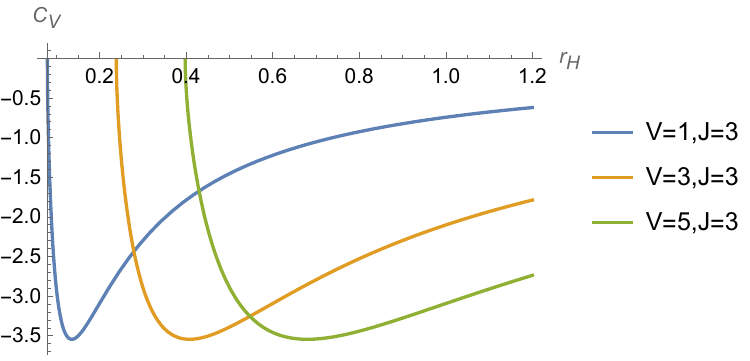}
\end{subfigure}
\caption{$d=4$, $C_V - r_H$ diagrams. (Branch A)}
\label{CV-rH-4-A}
\end{figure}

\begin{figure}[htbp]
\centering
\begin{subfigure}{0.4\textwidth}
    \includegraphics[width=\textwidth]{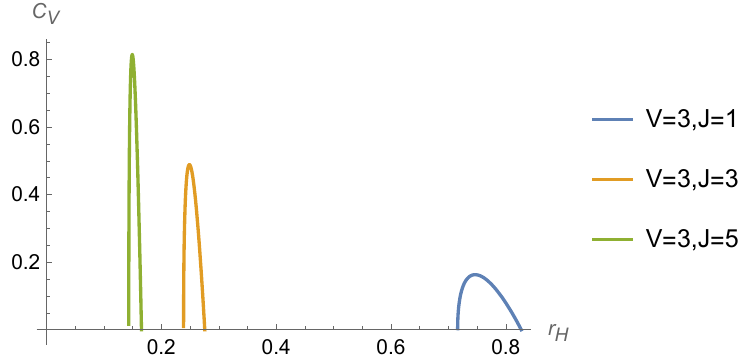}
\end{subfigure}
\begin{subfigure}{0.4\textwidth}
    \includegraphics[width=\textwidth]{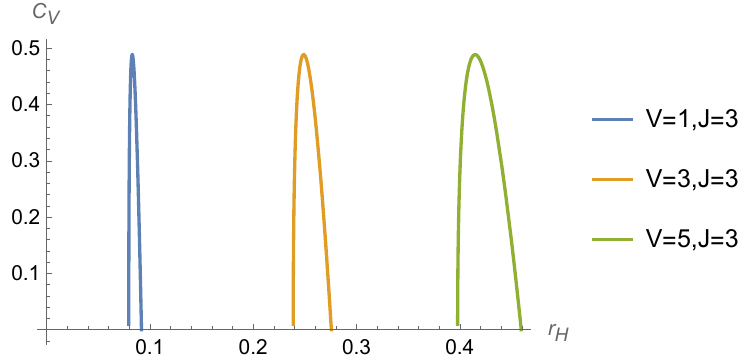}
\end{subfigure}

\caption{$d=4$, $C_V - r_H$. (Branch B) Only small positive regions remain.}
\label{CV-rH-4-B}
\end{figure}

For 4 dimensional cases, heat capacities as functions of $r_H$, and can be evaluated, the range of $r_H$ is chosen to meet $r_H > r_0$ to make sure horizon exists.
and the results are shown in Fig. \ref{CP-rH-4}, \ref{CV-rH-4-A} and \ref{CV-rH-4-B}.
This behavior suggests that, in the isobaric ensemble, the black hole admits a thermodynamically stable region at small horizon radii. As shown in the figure, 
the region of positive isobaric heat capacity shifts toward smaller $r_H$ with increasing pressure $P$. 

For Branch A, the results indicates that heat capacity is always negative, and when $J$ get larger, 
the minimum value of heat capacity becomes smaller. For Branch B, as shown in the figure, 
only small positive region is allowed in order to make sure horizon exists. In these regions, ultraspinning black holes are locally stable.


\subsection{Heat Capacity of 5-Dimensional Ultraspinning Black Holes}
\begin{figure}[htbp]
\centering
\begin{subfigure}{0.4\textwidth}
    \includegraphics[width=\textwidth]{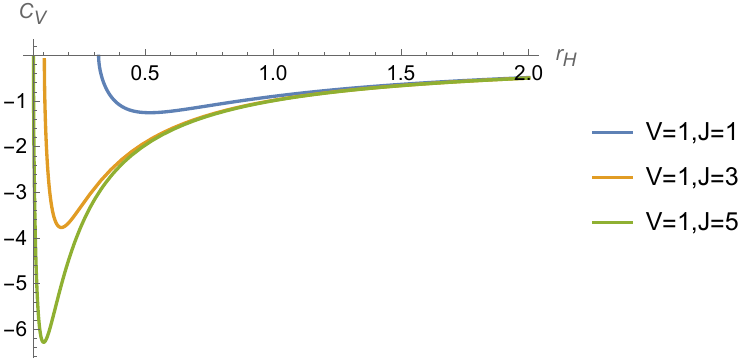}
\end{subfigure}
\begin{subfigure}{0.4\textwidth}
    \includegraphics[width=\textwidth]{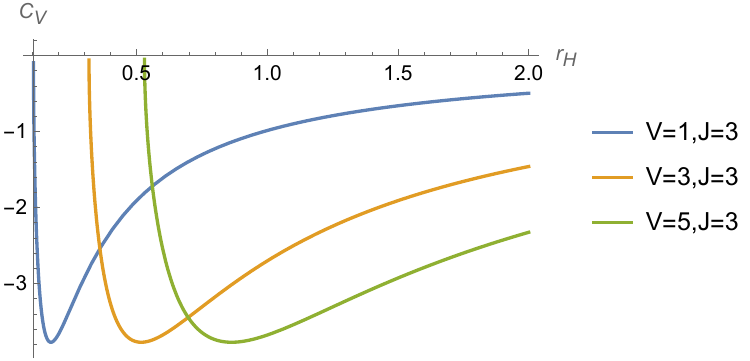}
\end{subfigure}\caption{$d=5$, $C_V - r_H$ diagram. (Branch A)}
\label{C_V - r_H - 5d - 1}
\end{figure}

\begin{figure}[htbp]
\centering
\begin{subfigure}{0.4\textwidth}
    \includegraphics[width=\textwidth]{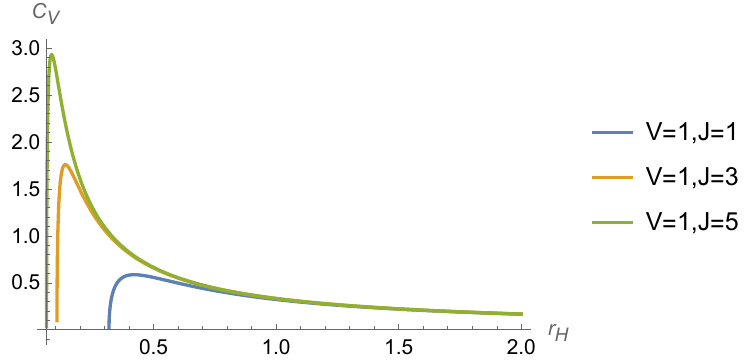}
\end{subfigure}
\begin{subfigure}{0.4\textwidth}
    \includegraphics[width=\textwidth]{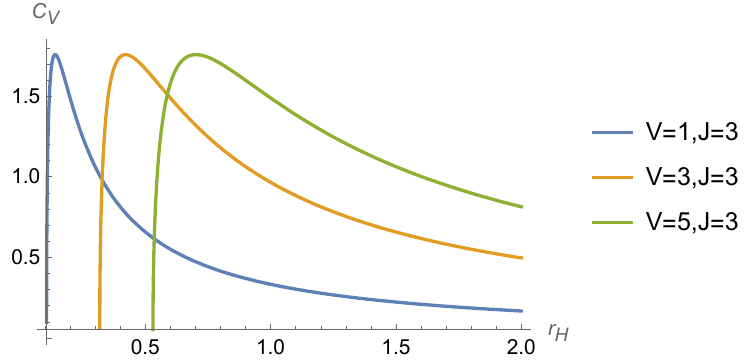}
\end{subfigure}\caption{$d=5$, $C_P - r_H$ diagram. (Branch B)}
\label{C_V - r_H - 5d - 2}
\end{figure}

For 5-dimensional case, the results of heat capacities are shown in Fig. \ref{C_P - r_H - 5d}, \ref{C_V - r_H - 5d - 1} and \ref{C_V - r_H - 5d - 2} :

\begin{figure}[htbp]
\centering
\begin{subfigure}{0.4\textwidth}
    \includegraphics[width=\textwidth]{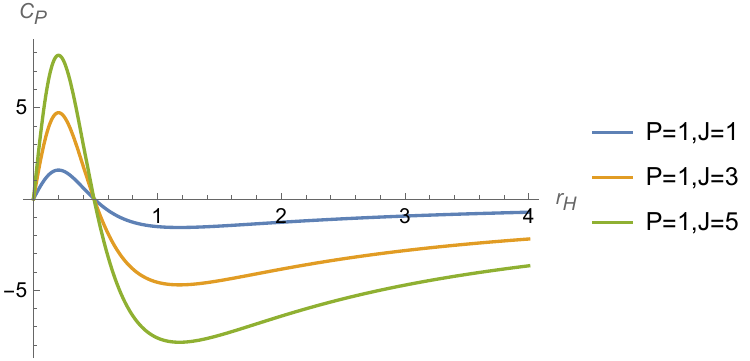}
\end{subfigure}
\begin{subfigure}{0.4\textwidth}
    \includegraphics[width=\textwidth]{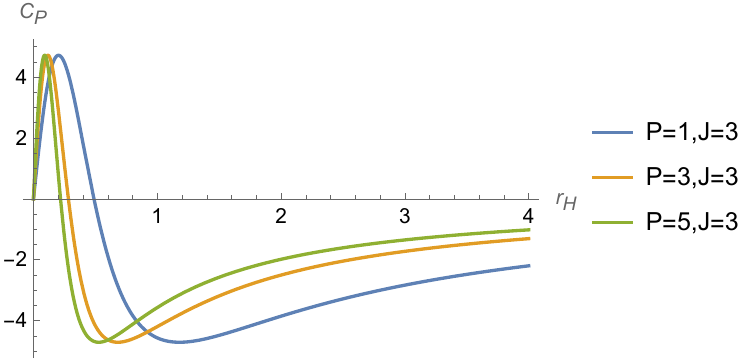}
\end{subfigure}\caption{$d=5$, $C_P - r_H$ diagram. }
\label{C_P - r_H - 5d}
\end{figure}
As shown in the figure Fig.~\ref{C_P - r_H - 5d}, the 5-dimensional case exhibits behavior similar to that of the 4-dimensional case, 
with a stable region characterized by a positive heat capacity appearing at small values of $r_H$.


The calculation of the heat capacity for the two branches in the constant-volume ensemble shows that, although two black holes with different parameters 
${l,\mu}$ may share identical thermodynamic quantities, their differences can be distinguished through their stability properties. 

For the case of smaller $l$
and larger $\mu$ (Branch A), the heat capacity is negative for black holes of all sizes, indicating that such configurations are thermodynamically unstable.
By contrast, for solutions with larger 
$l$ and smaller $\mu$ (Branch B), the heat capacity remains positive throughout the parameter space considered, implying that this class of solutions is thermodynamically stable.
\newpage
\subsection{Heat Capacity of 6 and Higher Dimensional Ultraspinning Black Holes}


The six-dimensional case exhibits qualitatively different behavior. 
For constant-volume cases, a critical point arises at a finite value of 
$r_H$, where the heat capacity diverges. To the left of this divergence, the heat capacity remains negative, indicating that sufficiently small black holes are thermodynamically unstable. To the right of the divergence, 
the heat capacity develops a positive region, implying local thermodynamic stability in this regime. However, for sufficiently large $r_H$, the heat capacity becomes negative again, rendering the black hole unstable once more.

For neutral black holes in higher dimensions, the behavior is analogous to that observed in six dimensions, differing only in quantitative details. 
Therefore, results are presented only up to the six-dimensional case in following discussion.

\begin{figure}[htbp]
\centering
\begin{subfigure}{0.4\textwidth}
    \includegraphics[width=\textwidth]{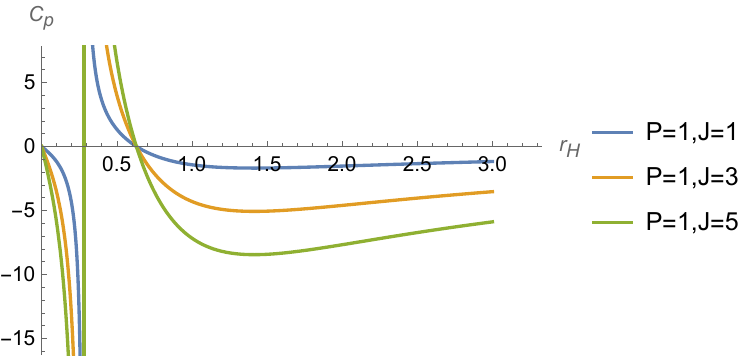}
\end{subfigure}
\begin{subfigure}{0.4\textwidth}
    \includegraphics[width=\textwidth]{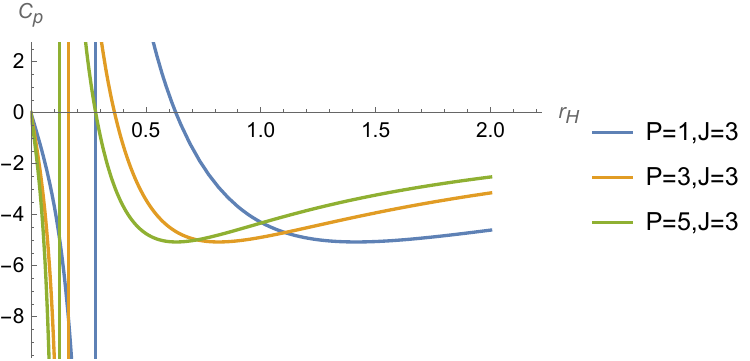}
\end{subfigure}\caption{$d=6$, $C_P - r_H$ diagram.}
\label{CP-rH-6}
\end{figure}

As indicated by Fig.~\ref{CP-rH-6}, the location of the critical point depends solely on the value of $P$. 
In terms of stability, a region with positive heat capacity exists to the right of the critical point.
\begin{figure}[htbp]
\centering
\begin{subfigure}{0.4\textwidth}
    \includegraphics[width=\textwidth]{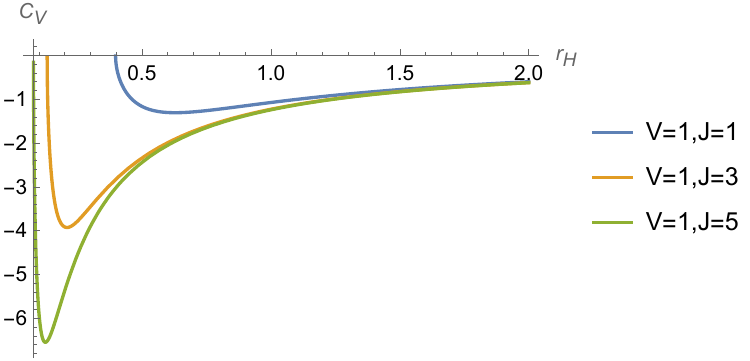}
\end{subfigure}
\begin{subfigure}{0.4\textwidth}
    \includegraphics[width=\textwidth]{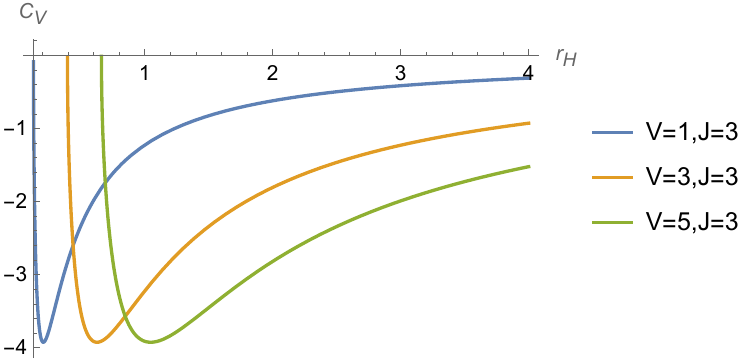}
\end{subfigure}\caption{$d=6$, $C_V - r_H$. (Branch A)}
\label{C_V - r_H - 6d - 1}
\end{figure}

\begin{figure}[htbp]
\centering
\begin{subfigure}{0.4\textwidth}
    \includegraphics[width=\textwidth]{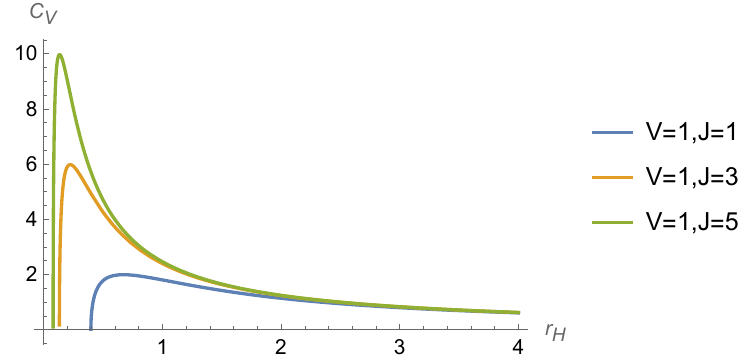}
\end{subfigure}
\begin{subfigure}{0.4\textwidth}
    \includegraphics[width=\textwidth]{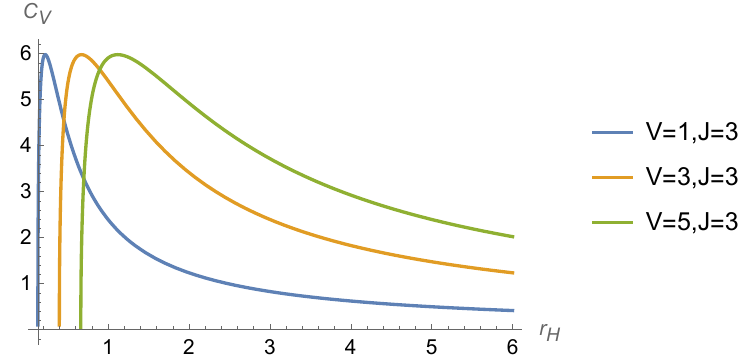}
\end{subfigure}\caption{$d=6$, $C_V - r_H$ diagram. (Branch B)}
\label{C_V - r_H - 6d - 2}
\end{figure}
The behavior of the heat capacity at constant pressure is similar to that in the five-dimensional case: 
black holes with smaller $l$ and larger $\mu$ (Branch A) are thermodynamically unstable, 
whereas the opposite regime (Branch B) corresponds to thermodynamically stable configurations.
\newpage
\subsection{Heat Capacity of 4 Dimensional Charged Ultraspinning Black Holes}



Similarly, the heat capacity of charged black holes can also be computed. 
As in the neutral case, the charged black holes also exhibits regions where the heat capacity is positive and negative; 
however, the distribution of these regions is more complex.

For the heat capacity at constant pressure, 
a given set of parameters $\{P, J, Q\}$ may correspond to three distinct solutions for $\{l, \mu, q\}$. 
Among these solutions, there are two branches with the value of $l$ in the same, while one branch is characterized by smaller $\mu$ and larger $q$, 
and the other exhibits the opposite behavior.
In addition, the range of $r_H$ is chosen to meet $r_H > r_0$ to make sure horizon exists.

However, in the third solution, 
the horizon radius $r_H$ is completely fixed 
by the values of ${V,J}$. 
In this case, the temperature $T$ 
cannot be treated as an independent thermodynamic state variable. 
As a result, the standard thermodynamic analysis 
based on independent variations of state variables is not applicable, 
and this solution will not be considered further.

\begin{figure}[htbp]
\centering
\begin{subfigure}{0.30 \textwidth}
    \centering
    \includegraphics[width=\textwidth]{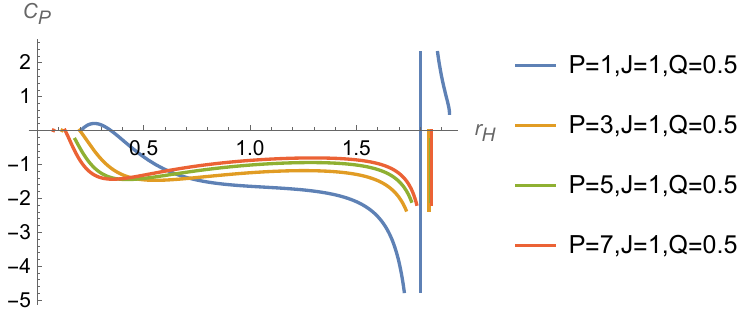}
\end{subfigure}
\begin{subfigure}{0.30 \textwidth}
    \centering
    \includegraphics[width=\textwidth]{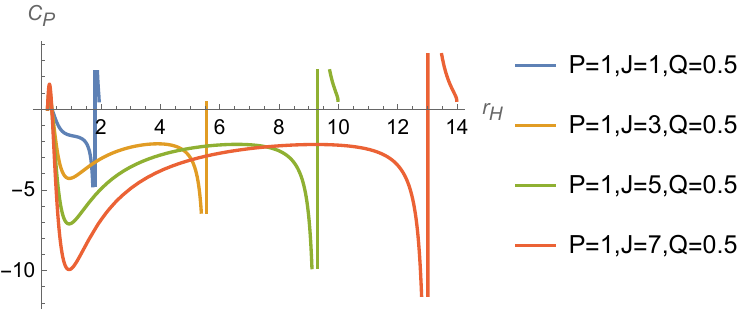}
\end{subfigure}
\begin{subfigure}{0.30 \textwidth}
    \centering
    \includegraphics[width=\textwidth]{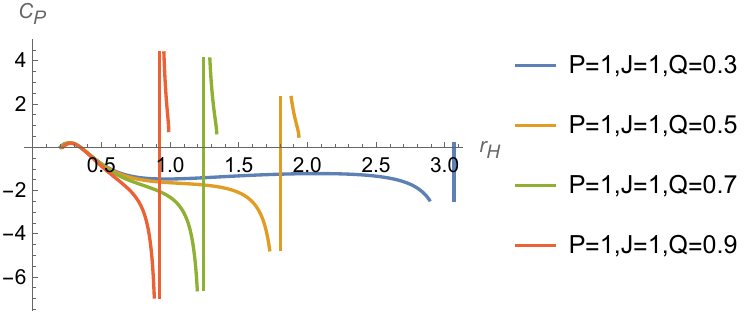}
\end{subfigure}

\caption{$d=4$, charged case, $C_P - r_H$ diagrams (Branch A). }
\label{CP-rh-4-Q-A}

\end{figure}

\begin{figure}[htbp]
\centering
\begin{subfigure}{0.30 \textwidth}
    \centering
    \includegraphics[width=\textwidth]{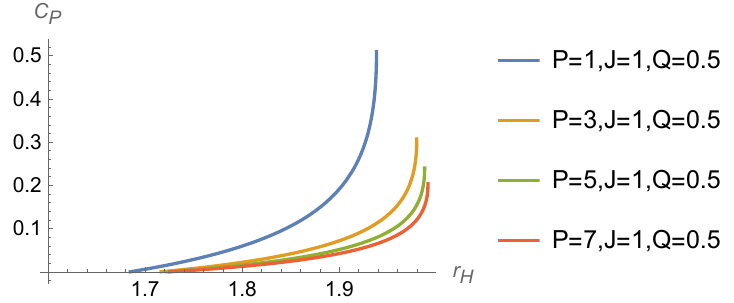}
\end{subfigure}
\begin{subfigure}{0.30 \textwidth}
    \centering
    \includegraphics[width=\textwidth]{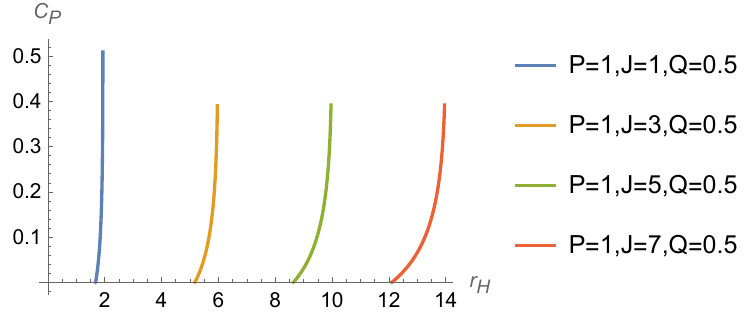}
\end{subfigure}
\begin{subfigure}{0.30 \textwidth}
    \centering
    \includegraphics[width=\textwidth]{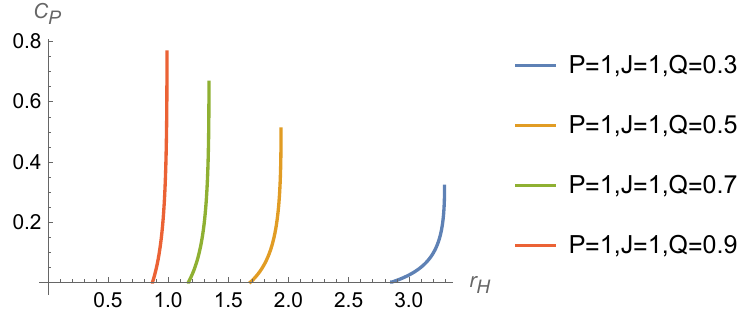}
\end{subfigure}

\caption{$d=4$, charged case, $C_P - r_H$ diagrams (Branch B).}
\label{CP-rh-4-Q-B}

\end{figure}

The result of $C_P$ is shown in Fig.~\ref{CP-rh-4-Q-A} and \ref{CP-rh-4-Q-B}.
For Branch A, the diagrams indicate that, when the heat capacity at constant pressure is considered, 
depending on choice of thermodynamic quantities, 
stable regions may appear on right side (e.g. $P = 5, J = 1, Q = 0.5$) or both sides (e.g. $P = 1, J = 1, Q = 0.5$) of the critical point.
For Branch B, the diagrams indicate that, 
when the heat capacity at constant pressure is considered, 
one stable regions is allowed in order to make sure horizon exists. 
Moreover, the heat capacity does not increase indefinitely. When $r_H$ becomes large enough, $C_P$ do not exist because 
there is no black hole solution associated with given $P$ and $J$.

\begin{figure}[htbp]
\centering
\begin{subfigure}{0.30 \textwidth}
    \centering
    \includegraphics[width=\textwidth]{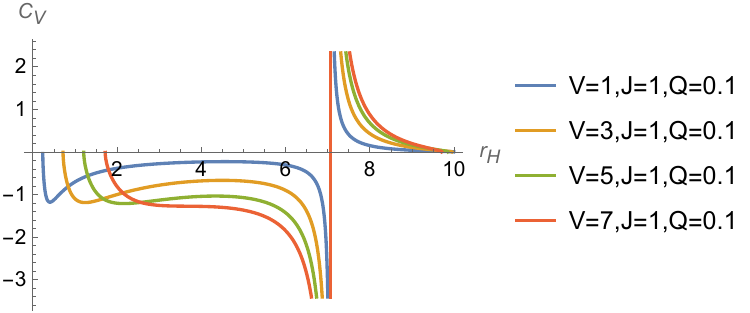}
\end{subfigure}
\begin{subfigure}{0.30 \textwidth}
    \centering
    \includegraphics[width=\textwidth]{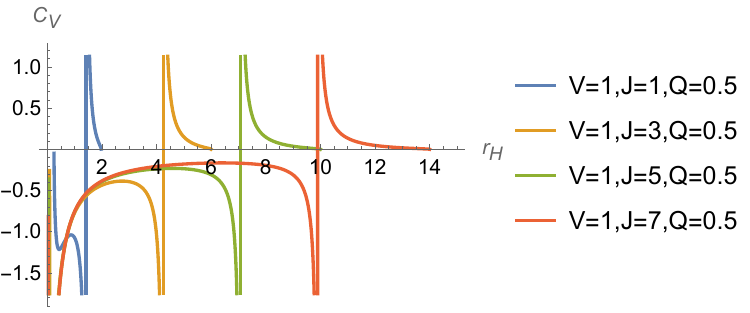}
\end{subfigure}
\begin{subfigure}{0.30 \textwidth}
    \centering
    \includegraphics[width=\textwidth]{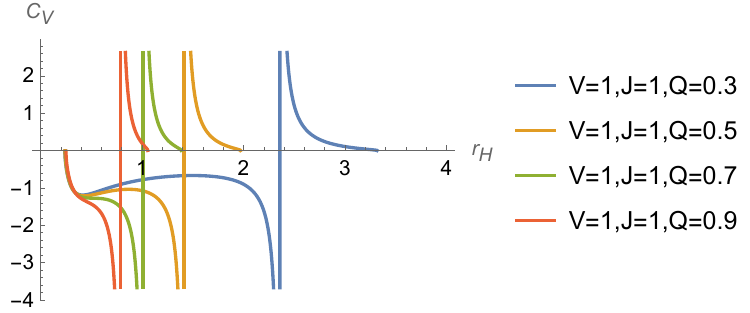}
\end{subfigure}

\caption{$d=4$, charged cases, $C_V - r_H$ diagrams (Branch A).}

\label{CV-rH-4-Q-A}
\end{figure}

\begin{figure}[htbp]
\centering
\begin{subfigure}{0.30 \textwidth}
    \centering
    \includegraphics[width=\textwidth]{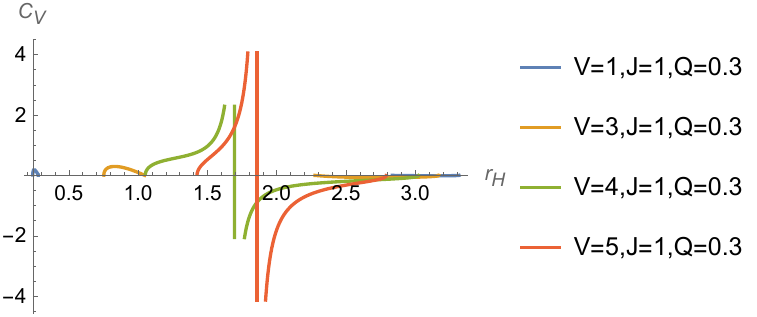}
\end{subfigure}
\begin{subfigure}{0.30 \textwidth}
    \centering
    \includegraphics[width=\textwidth]{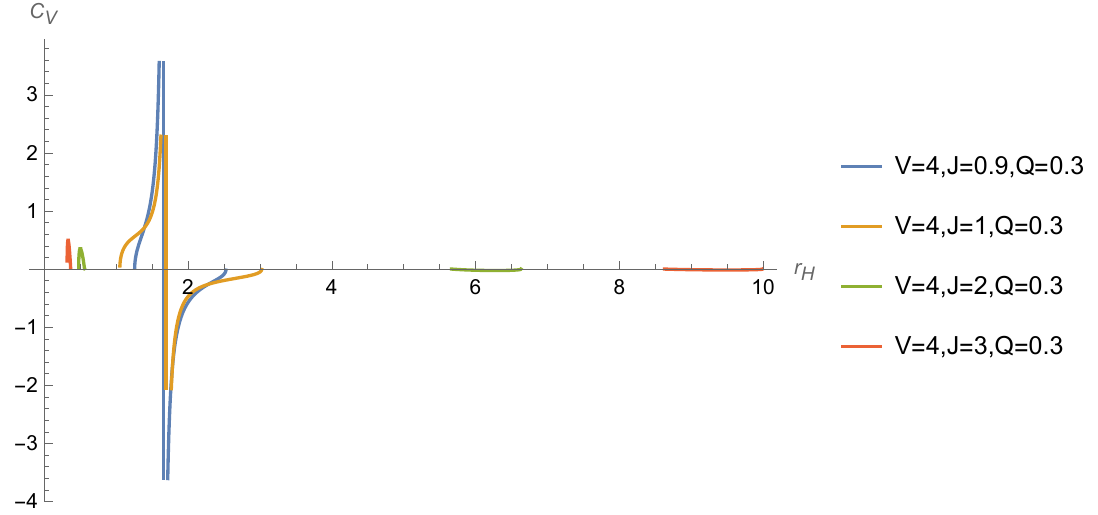}
\end{subfigure}
\begin{subfigure}{0.30 \textwidth}
    \centering
    \includegraphics[width=\textwidth]{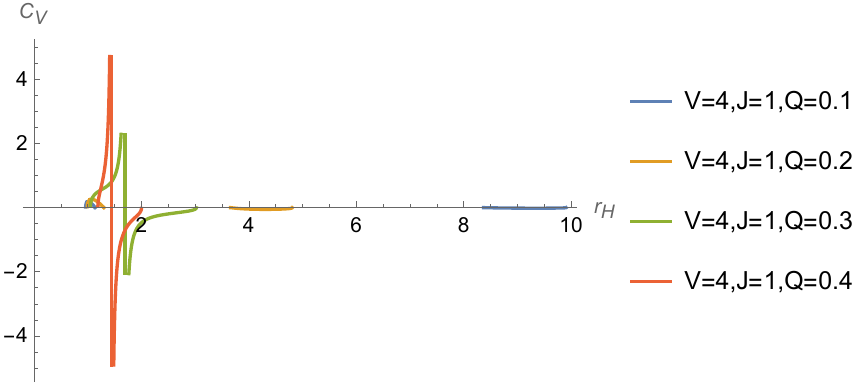}
\end{subfigure}

\caption{$d=4$, $C_V - r_H$ diagrams for charged case (Branch B).}
\label{CV-rH-4-Q-B}
\end{figure}

The results for $C_V$ for charged cases are shown in Fig.~\ref{CV-rH-4-Q-A} and \ref{CV-rH-4-Q-B}.
For Branch A, a stable region exists to the right of the critical point, 
and the location of the critical point is independent of the value of 
$V$.
For Branch B, the diagrams indicates that when considering the heat capacity at constant volume, 
there are two cases for permit range of $r_H$. As shown in \ref{CV-rH-4-Q-B}, when volume $V$ increases, 
the range of $r_H$ changes from two separate regions to one connected region, each containing a locally stable region with positive heat capacity.
However, when $V$ becomes too large, all values of $r_H$ are ruled out by the existence of horizon. Similar behavior occur when angular momentum decreases or charge $Q$ increases.

\newpage
\subsection{Brief Summary of Heat Capacity}
Here, we briefly summarize the heat capacity in different case in Table \ref{tab:capacity}. 
As shown in the table, both $C_P$ and $C_V$ 
can exhibit highly nontrivial branch-dependent behavior, 
with sign changes and stable intervals appearing in different parameter regimes, 
which represents several locally thermodynamic stable regions for ultraspinning black holes.

\begin{table}[htbp]
\centering
\renewcommand{\arraystretch}{1.25}
\begin{tabular}{|c|c|c|c|}
\hline
Cases
& Isobaric Heat Capacity $C_P$ 
& Isochoric Heat Capacity $C_V$ 
& Summary\\
\hline
\makecell{$d=4$, \\neutral}
&
\makecell[l]{%
$C_P>0$ for small $r_H$;\\
$C_P<0$ for large $r_H$}
&
\makecell[l]{%
Branch A ($l$ small, $\mu$ large): \\
always negative;\\
Branch B ($l$ large, $\mu$ small):\\
\,\, positive region near extremality}
&
\makecell[l]{%
Local stability appears\\
in the isobaric ensemble\\
and isochoric ensemble\\
for specific branches}
\\
\hline
\makecell{$d=5$, \\neutral}
&
\makecell[l]{%
$C_P>0$ for small $r_H$;\\
$C_P<0$ for large $r_H$}
&
\makecell[l]{%
Branch A ($l$ small, $\mu$ large): \\always negative;\\
Branch B ($l$ large, $\mu$ small): \\always positive}
&
\makecell[l]{%
Stable regions exist;\\
stability is branch-dependent}
\\
\hline
\makecell{$d\geq 6$, \\neutral}
&
\makecell[l]{%
Critical point present;\\
$C_P>0$ to the right\\ of the critical point}
&
\makecell[l]{%
Branch A ($l$ small, $\mu$ large): \\always negative;\\
Branch B ($l$ large, $\mu$ small): \\always positive}
&
\makecell[l]{%
Critical behavior emerges\\
in higher dimensions}
\\
\hline
\makecell{$d=4$, \\charged}
&
\makecell[l]{%
Branch A ($q$ small, $\mu$ large):\\
\quad positive regions \\ on both sides of the critical point;\\
Branch B ($q$ large, $\mu$ small):\\
\quad one stable region}
&
\makecell[l]{%
Branch A (larger $\mu$, smaller $l,q$):\\
\quad single critical point;\\
\quad unstable on the left, \\
\quad stable on the right;\\
Branch B (smaller $\mu$, larger $l,q$):\\
\quad multiple allowed regions\\
\quad $C_V > 0$ for approciate parameters}
&
\makecell[l]{%
Electric charge enhances\\
the complexity \\ of the stability structure}
\\
\hline
\end{tabular}
\caption{Thermodynamic stability of ultraspinning black holes in different dimensions and charge configurations
Branch A has larger $\mu$ and Branch B is opposite.}
\label{tab:capacity}
\end{table}

\section{Revised Reverse Isoperimetric Inequality (RRII) and Its Constraints on Black Holes}

\subsection{Reverse isoperimetric inequality and Its Counter Example}
In Euclidean geometry, there is a famous inequality called "isoperimetric inequality":
\begin{equation}
    \mathcal{R} := \left[\frac{(d-1) V}{\omega_{d-2}}\right]^{\frac{1}{d-1}} \left(\frac{\omega_{d-2}}{A}\right)^{\frac{1}{d-2}} \leqslant  1
\end{equation}
in which
\begin{equation}
    \omega_d = \frac{2 \pi^{\frac{d+1}{2}}}{\Gamma\left(\frac{d+1}{2}\right)},
\end{equation}
is the volume of $d$ dimensional unit sphere. The equality is attained when the object is a sphere. 
The inequality can be interpreted as stating that, among all geometric objects with the same volume, a sphere has the smallest surface area.

In black hole thermodynamics, the thermodynamic volume and geometric volume can be defined, and in \cite{Pope.Gibbons.RII.PRD}, authors proposed a reverse isoperimetric inequality
should hold for all black holes, which means among all black holes with same volume, the Schwarzschild - AdS black hole has the largest surface area.

The specific inequality is: 
\begin{equation}
\mathcal{R} := \left[\frac{(d-1) V}{\omega_{d-2}}\right]^{\frac{1}{d-1}} \left(\frac{\omega_{d-2}}{A}\right)^{\frac{1}{d-2}} \geqslant 1, 
\end{equation}
and the equality is attained for Schwarzschild - AdS black hole.

Although the expectation of \cite{Pope.Gibbons.RII.PRD} is to provide a criterion 
to choose proper volume of a black hole and it does hold for a lot of cases, 
in \cite{PhysRevLett.115.031101.Mann.Hennigar.2015}, the authors noticed the ultraspinning black holes violate RII.

Since the period of coordinate $\psi$ is compacted as $\mu$, the result of evaluating the area of 2 - dimensional sphere is $\omega_2 = 2 \mu$, 
and the volume of d - dimensional ball is $\omega_d = \frac{ \mu \pi^{\frac{d-1}{2}}}{\Gamma\left(\frac{d+1}{2}\right)}$,
and the ratio is: 
\begin{equation}
    \mathcal{R} = \left(\frac{r_H^2}{r_H^2 + l^2}\right)^{\frac{1}{(d-1)(d-2)}} < 1.
\end{equation}
which shows ultraspinning black holes in any dimension violates RII.

\subsection{Revised Reverse Isoperimetric Inequality(RRII)}

It should be emphasized that the revised reverse isoperimetric inequality 
provides a geometric constraint on the entropy–volume relation, 
while thermodynamic stability in this work 
is diagnosed via heat capacities. 
These two notions are logically independent, 
and will be analyzed separately below.

In \cite{PhysRevLett.131.241401.Amo.2023}, authors proposed a revised reverse isoperimetric inequality(RRII), 
which contains mass ($M$), angular momentum ($J_i$) and volume ($V$):
\begin{equation}
    A(M,J_i,V) \leqslant A_{Kerr}(M,J_i,V),
\end{equation}
in which $A_{Kerr}(M,J_i,V)$ represents the area of Kerr - AdS black holes with same $M$, $J_i$ and $V$.

Since ultraspinning violates RII, it is worth checking whether ultraspinning black holes also violates RRII. 
And it turns out that if RRII is treated as a preset condition,
 it can be used to constrain the range of parameters in ultraspinning black holes.

\begin{figure}[!htbp] 
    \centering
    \begin{subfigure}{\textwidth}
        \centering
        \includegraphics[width=0.4\textwidth]{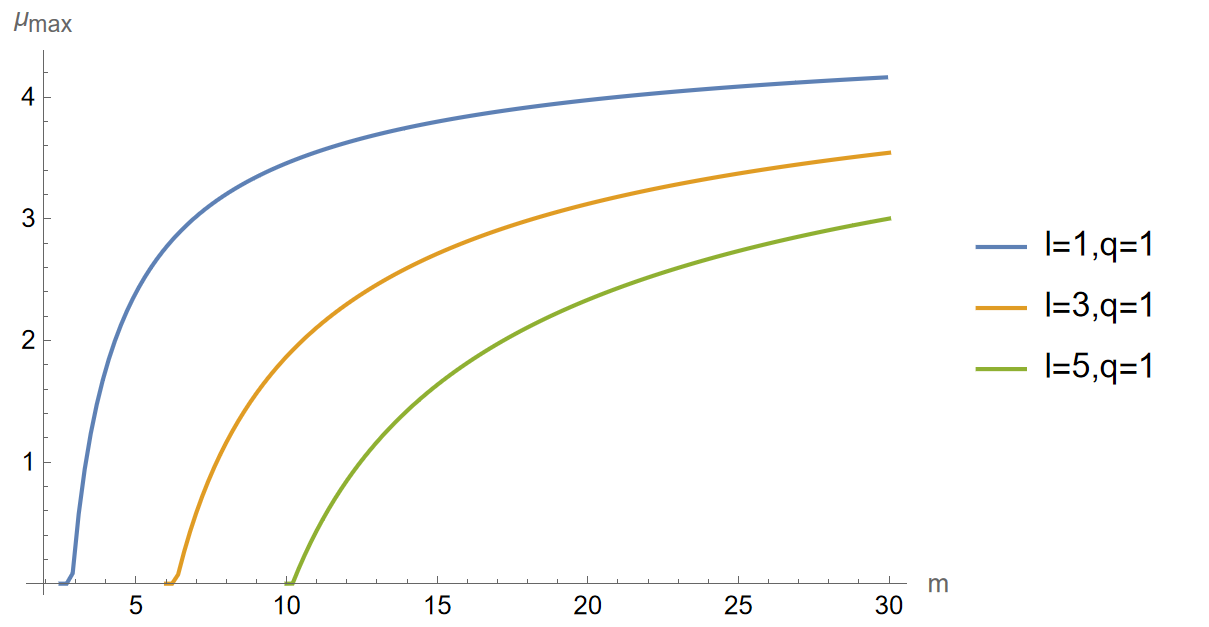}
        \includegraphics[width=0.4\textwidth]{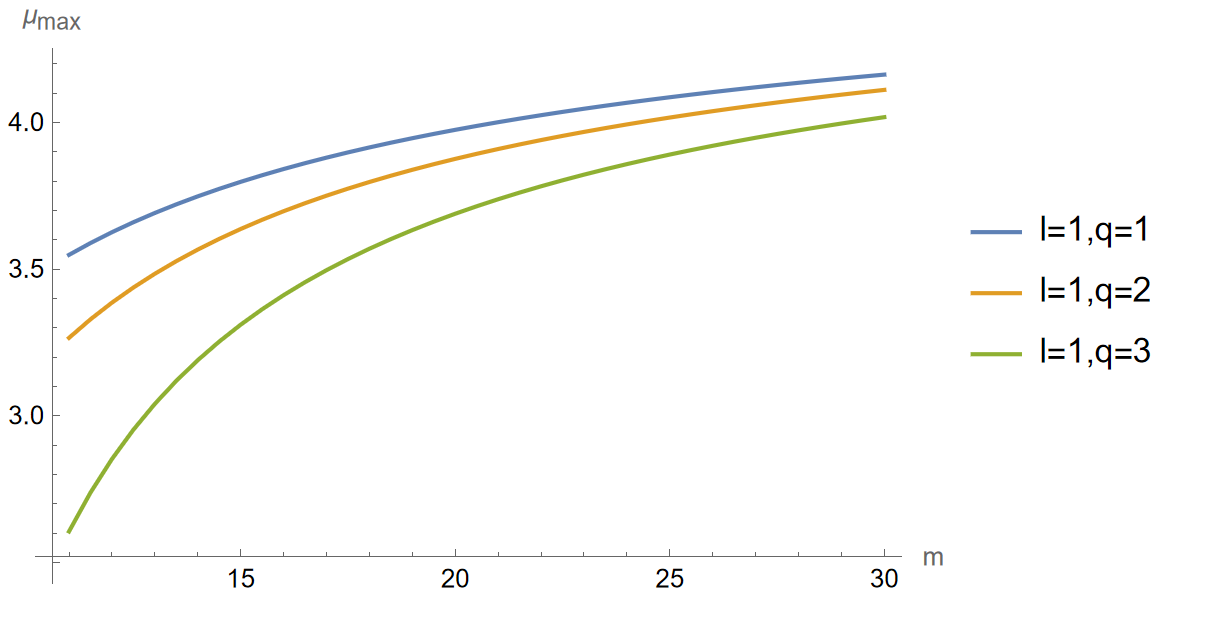}
        \caption{$\mu_{max}$-$m$, $l$=1, $q$=1.}
        \label{subfig:l}
    \end{subfigure}
    
    \vspace{1cm} 
    
    \begin{subfigure}{\textwidth}
        \centering

        \includegraphics[width=0.4\textwidth]{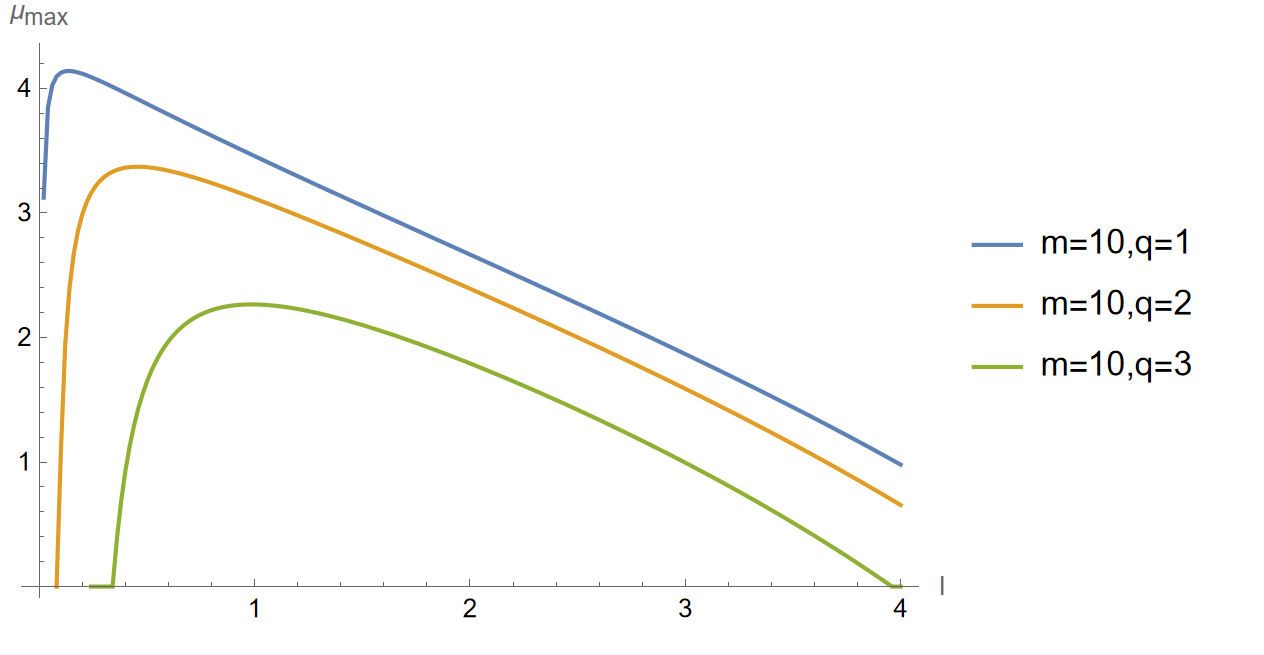}
        \includegraphics[width=0.4\textwidth]{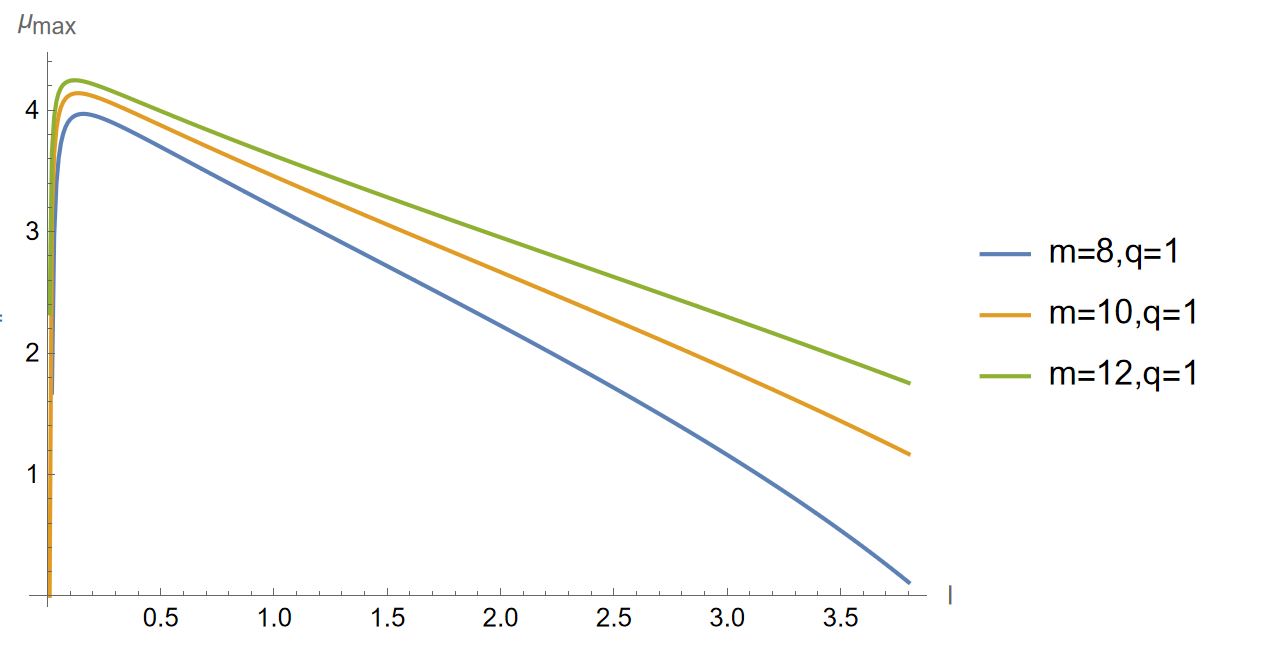}
        \caption{$\mu_{max}$-$l$, $l$=1, $q$=1.}
        \label{subfig:m}
    \end{subfigure}
    
    \vspace{1cm} 
    
    \begin{subfigure}{\textwidth}
        \centering
        \includegraphics[width=0.4\textwidth]{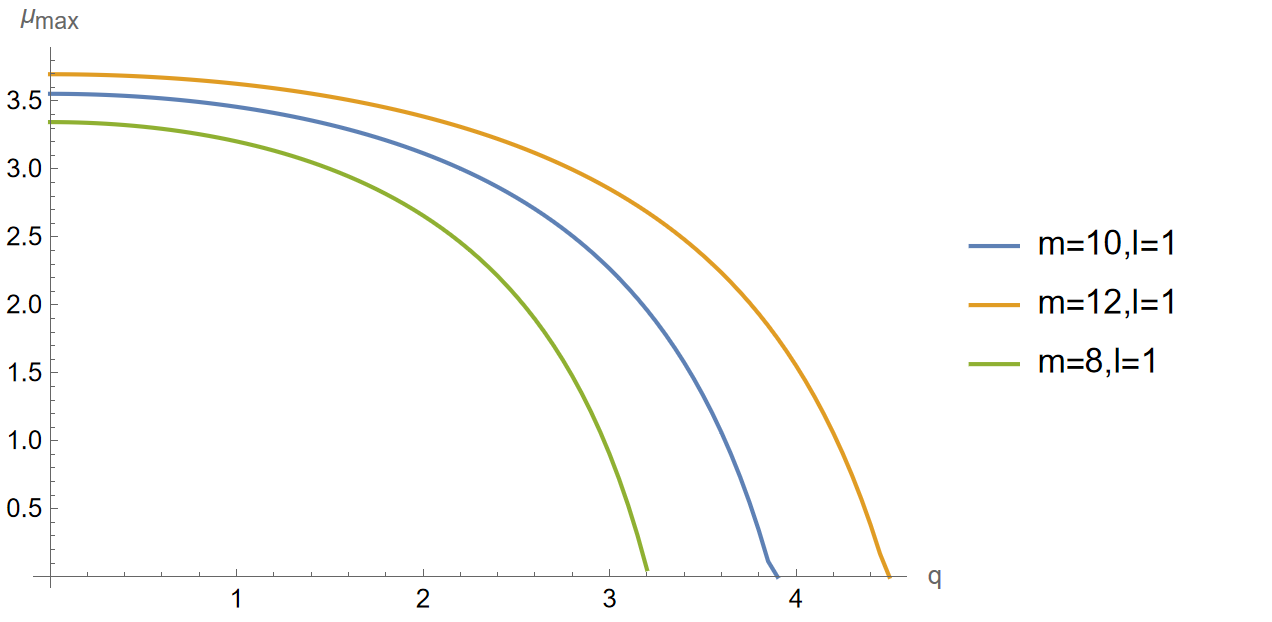}
        \includegraphics[width=0.4\textwidth]{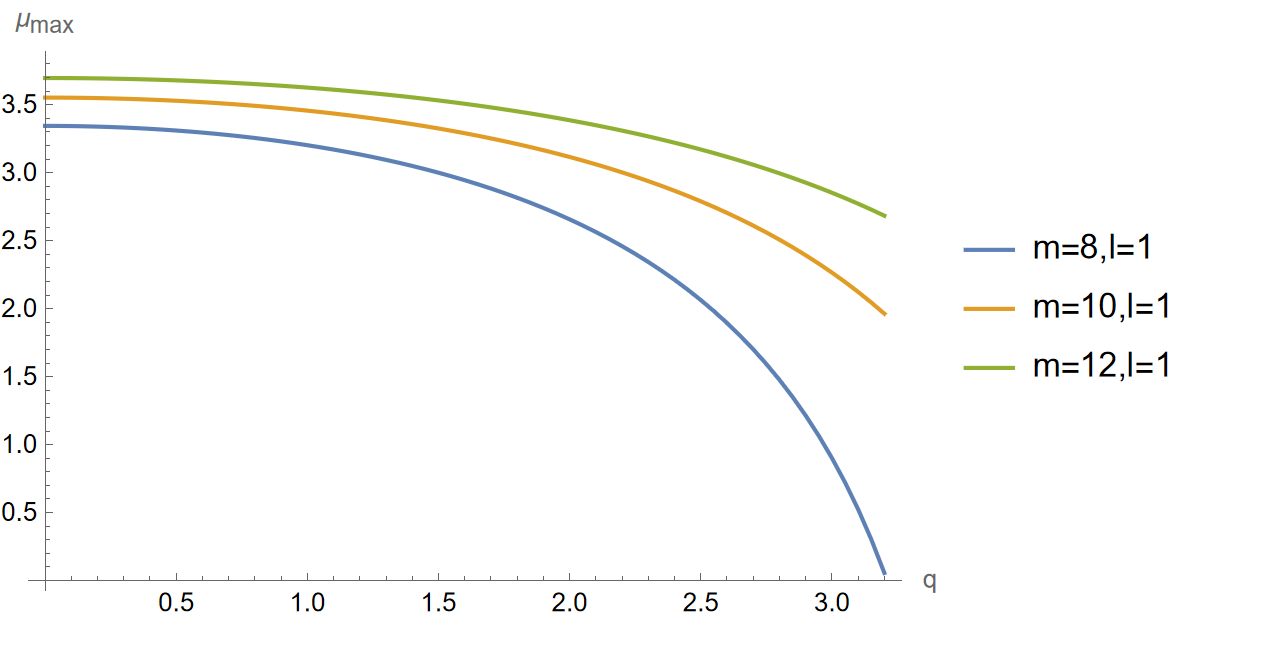}
        \caption{$\mu_{max}$-$q$, $m$=10, $l$=1.}
        \label{subfig:q}
    \end{subfigure}
      
    \caption{$\mu_{max}$ constraints for different parameters}
    \label{fig:参数取值对mu的限制}
\end{figure}

Some observation from Fig.\ref{fig:参数取值对mu的限制}:

(1)RRII does constrain the maximum of $\mu$;

(2)the results shows the constrain on $m, q, l$. For $m$, there is a strengthened lower bound than it given by existence of root of $\Delta(r)$. 
For $l$ and $q$, there exist upper bound.









\newpage
\section{Summary}



In this paper, 
a systematic investigation of the thermodynamics and stability properties of ultraspinning black holes has been presented. 
The central new result is a comprehensive analysis of their heat capacities. 
By examining different thermodynamic ensembles and solution branches, 
it is shown that ultraspinning black holes are not generically thermodynamically unstable. 
Instead, stable regions characterized by positive heat capacity can exist in certain parameter regimes, 
providing explicit counterexamples to the conjecture that super-entropic black holes must always be unstable.

The thermodynamic quantities and the first law of ultraspinning black holes were re-derived within the Iyer–Wald formalism, 
supplemented by the integrability condition for conserved charges. 
Although the resulting expressions coincide with those previously obtained using the conformal completion approach, 
this agreement constitutes a nontrivial consistency check of the applicability 
of the Iyer–Wald framework to black holes with non-compact horizon topology.

The constraints imposed by the (revised) reverse isoperimetric inequality were also examined. 
While ultraspinning black holes violate the original reverse isoperimetric inequality, 
the revised version remains applicable and yields nontrivial bounds on the parameter 
$\mu$. To avoid obscuring the main discussion of thermodynamic stability, 
the detailed analysis of this issue has been relegated to an appendix.

Several directions for future work naturally arise. 
In particular, a more systematic understanding of 
the relation between the reverse isoperimetric inequality and thermodynamic stability remains to be developed. 
Moreover, it would be interesting to explore how thermodynamic stability properties are reflected in observable quantities, 
such as black hole shadows or other phenomenological signatures.

\newpage
\begin{acknowledgments}
    ZD wants to thank Sijie Gao, Yunjiao Gao, Shun Jiang and Zhe Meng for helpful discussion.
\end{acknowledgments}

\appendix

\section{First Law of Black Holes with Cosmological Constant}
    \subsection{Variation of Lagrangian}
    The Einstein-Hillbert-Maxwell action with a cosmological constant can be written as:
        \begin{equation}
            \boldsymbol{L} = \frac{1}{16 \pi}(R - 2 \Lambda - F^{ab}F_{ab})\boldsymbol{\epsilon},
            \label{LEHM_A}
        \end{equation}
    in which the bold symbols represent differential forms, $R$ is the Ricci scalar, $\boldsymbol{F} = d \boldsymbol{A}$ is the electromagnetic field strength tensor,
    $\boldsymbol{\epsilon}$ is the associated volume form, which satisfies $\nabla_a \boldsymbol{\epsilon}=0$, 
    $\Lambda$ is the cosmological constant.
    
    By variation, one can deduce the equation of motion and boundary terms.
    These boundary terms will be used in constructing Noether flow and Noether charge.
    Different thermodynamic quantities in black hole thermodynamics can be viewed as Noether charges associated with different Killing vectors.
    
    By variation of $\boldsymbol{L}$:
    $$\delta \boldsymbol{L} = \boldsymbol{E}_{g} \delta g^{ab} + \boldsymbol{E}_{A} \delta A_a - \frac{1}{8\pi} \boldsymbol{\epsilon}\delta \Lambda + d\boldsymbol{\Theta},$$
    in which all $\boldsymbol{E}$ represent tensors associated to EOMs, $d\Theta$ is the so-called boundary term, which will be used to construct Noether charge, and $\Lambda$ is also viewed as a variable.
    In addition, the specific expressions of $\boldsymbol{E}$ and $\boldsymbol{\Theta}$ is:
    \begin{equation}
        \boldsymbol{E}_g = \frac{1}{16\pi} \left(R_{ab} - \frac{1}{2}R g_{ab} + \Lambda g_{ab} - 8\pi T^{EM}_{ab}\right) \boldsymbol{\epsilon},
    \end{equation}
    \begin{equation}
        \boldsymbol{E}_{A} = \frac{1}{4\pi} \nabla_a F^{ab} \boldsymbol{\epsilon},
    \end{equation}
    \begin{equation}
        \boldsymbol{\Theta}(\Psi,\delta\Psi) = \frac{1}{16 \pi} \left[g^{ce} g^{ab}\left(\nabla_b \delta g_{ac} - \nabla_c \delta g_{ab}\right) - 4 F^{eb} \delta A_b\right] \epsilon_{elmn}.
    \end{equation}
    Here $\Psi$ collectively denotes $\{g_{ab}, A_a\}$, and the electromagnetic stress-energy tensor $T^{EM}_{ab}$ is given by:
    \begin{equation}
        T^{EM}_{ab} = -\frac{1}{16 \pi} g_{ab} F_{cd}F^{cd} + \frac{1}{4\pi} F_a^{\ \:c}F_{bc}.
    \end{equation}
    \subsection{Construction of Noether Charge}
    After evaluating boundary term from Lagrangian, one can construct Noether current from it:
    \begin{equation}
        \boldsymbol{J} = \boldsymbol{\Theta}(\psi,\mathcal{L}_{\xi}\psi) - \xi \cdot \boldsymbol{L}
        \label{J_A}
    \end{equation}
    Here $\xi^a$ is associated with diffeomorphism invariance.
    It can be proven that $\boldsymbol{J}$ can always be decomposed into\cite{PhysRevD.52.4430.Wald.1995}:
    $$\boldsymbol{J} = \boldsymbol{C}_{\xi} - d \boldsymbol{Q},$$
    in which, $\boldsymbol{C}[\xi] = 0$ when all equations of motion hold.For our case \eqref{LEHM}, $\boldsymbol{C}[\xi]$ and $\boldsymbol{Q}[\xi]$ can be evaluated as 
    (a convenient way to obtain this form is to take the Hodge dual of \eqref{J_A}, separate it into a total divergence and terms that vanish when the equations of motion hold, and then take the Hodge dual again):
    \begin{equation}
        \boldsymbol{C}_\xi = \frac{1}{8\pi}\xi^c \left[R_{dc} - \frac{1}{2}R g_{dc} + \Lambda g_{dc} - 8\pi T^{EM}_{dc}\right]\boldsymbol{\epsilon}^d_{\ lmn} + \frac{1}{4\pi} A_a \xi^a \nabla_b F^{\ \:b}_d \boldsymbol{\epsilon}^d_{\ lmn},
    \end{equation}
    \begin{equation}
        \boldsymbol{Q}_\xi = -\frac{1}{16\pi} \left(\nabla^d \xi^b + 2 A_a \xi^a F^{db} \right) \boldsymbol{\epsilon}_{mndb}.
        \label{Q_A}
    \end{equation}

    Define a new variation operator called $\bar{\delta}$,which means not to variate $\xi^a$ and the relation of $\delta$ and $\bar{\delta}$ is $\bar{\delta} \boldsymbol{X}_{\xi} = \delta \boldsymbol{X}_{\xi} - \boldsymbol{X}_{\delta \xi}$.
    Then the variation of $\boldsymbol{J}$ is:
    \begin{equation}
        \begin{aligned}
            \bar{\delta} \boldsymbol{J} & = \bar{\delta} \boldsymbol{\Theta} - \xi \cdot  \bar{\delta} \boldsymbol{L}\\
            & = \bar{\delta} \boldsymbol{\Theta} - \xi \cdot \delta \boldsymbol{L}\\
            & = \bar{\delta} \boldsymbol{\Theta} - \xi \cdot \boldsymbol{E}_{\psi} \delta \psi - \xi \cdot d \boldsymbol{\Theta}(\psi,\delta \psi) + \frac{1}{8 \pi}\xi \cdot \boldsymbol{\epsilon} \delta \Lambda\\
            & = \bar{\delta} \boldsymbol{\Theta} - \xi \cdot \boldsymbol{E}_{\psi} \delta \psi + d(\xi \cdot \boldsymbol{\Theta}) - \mathcal{L}_{\xi} \boldsymbol{\Theta} + \frac{1}{8 \pi}\xi \cdot \boldsymbol{\epsilon} \delta \Lambda\\
            & = \boldsymbol{\omega}(\psi,\bar{\delta}\psi,\mathcal{L}_{\xi}\psi) - \xi \cdot \boldsymbol{E}_{\psi}\delta \psi  +  d(\xi \cdot \Theta) + \frac{1}{8 \pi}\xi \cdot \boldsymbol{\epsilon} \delta \Lambda.
        \end{aligned}
    \end{equation}
    when $\xi$ in above formula is a Killing vector field and represents a symmetry of dynamical fields,i.e. $\mathcal{L}_{\xi}\psi = 0$, $\omega(\psi,\bar{\delta}\psi,\mathcal{L}_{\xi}\psi) = 0$ \cite{PhysRevD.50.846.Wald94}.
    Then
    \begin{equation}
        \bar{\delta} \boldsymbol{J} =  \bar{\delta} \boldsymbol{C}_\xi + d( \bar{\delta} \boldsymbol{Q}_\xi) = - \xi \cdot \boldsymbol{E}_{\psi}\delta \psi  +  d(\xi \cdot \Theta) + \frac{1}{8 \pi}\xi \cdot \boldsymbol{\epsilon} \delta \Lambda.
    \end{equation}
    When EOMs hold,$\boldsymbol{C}_\xi =0$,
    \begin{equation}
        d(\bar{\delta}\boldsymbol{Q}_\xi - \xi \cdot \boldsymbol{\Theta}) = \frac{1}{8\pi} \xi \cdot \boldsymbol{\epsilon}\delta \Lambda.
    \end{equation}
    
    Integrate both sides from horizon to any surface of $t=const$ and $r=const$ living out of horizon, called $S$, then apply Stokes' theorem to LHS and separate RHS into two parts:
    \begin{equation}
        \int_S \bar{\delta}\boldsymbol{Q}_\xi - \xi \cdot \boldsymbol{\Theta} - \int_H \bar{\delta}\boldsymbol{Q}_\xi - \xi \cdot \boldsymbol{\Theta} = \int_{r=0}^{r=r_s} \frac{1}{8\pi} \xi \cdot \boldsymbol{\epsilon}\delta \Lambda - \int_{r=0}^{r=r_H} \frac{1}{8\pi} \xi \cdot \boldsymbol{\epsilon}\delta \Lambda.
    \end{equation}

    From the above equation, it is straightforward to see there exists a quantity is independent of the choice of integration surface:
    \begin{equation}
        \int_S (\bar{\delta}\boldsymbol{Q}_\xi - \xi \cdot \boldsymbol{\Theta}) - \int_{r=0}^{r=r_s} \frac{1}{8\pi} \xi \cdot \boldsymbol{\epsilon} \delta \Lambda.
    \end{equation} 
    When the Killing vector is chosen to be $t^a \equiv \left(\pdv{}{t}\right)^a$, the corresponding quantity is defined as variation of mass:
    \begin{equation}
        \delta M = \int_S \bar{\delta}\boldsymbol{Q}_t - t \cdot \boldsymbol{\Theta} - \int_{r=0}^{r=r_s} \frac{1}{8\pi} t \cdot \boldsymbol{\epsilon} \delta \Lambda.
        \label{dM}
    \end{equation}
    When the Killing vector is chosen to be $-\phi^a \equiv -\left(\pdv{}{\phi}\right)^a$, the corresponding quantity is defined as variation of angular momentum:
    \begin{equation}
        \begin{aligned}
            \delta J & = - \int_S \bar{\delta}\boldsymbol{Q}_\phi - \phi \cdot \boldsymbol{\Theta} - \int_{r=0}^{r=r_s} \frac{1}{8\pi} \phi \cdot \boldsymbol{\epsilon} \delta \Lambda.\\
                & = - \int_S \bar{\delta}\boldsymbol{Q}_\phi.
        \end{aligned}
    \end{equation}
    The second line is due to the integration after $\phi \cdot \boldsymbol{\Theta}$ and $\phi \cdot \epsilon$ vanish in $\phi$ direction.

    Fixing gauge condition as $\delta t^a = 0$ and $ \delta \phi^a = 0$. Under this,$\delta \boldsymbol{Q} = \bar{\delta} \boldsymbol{Q}$. 
    It then follows directly that:
    \begin{equation}
        J = -\int_S \boldsymbol{Q}_\phi
    \end{equation}
    from the definition of $\delta J$.

    Consider Killing vector $\xi^a = t^a + \Omega_H \phi^a$($\Omega:= - \frac{g_{t \phi}}{g_{\phi \phi}} |_H$,angular velocity of black hole).
    Then it can be organized from the definition of $\delta J$ and $\delta M$ that:
    \begin{equation}
        \delta M - \Omega_H \delta J = \int_H (\bar{\delta} \boldsymbol{Q}_\xi - \xi \cdot \Theta) - \int_{r=0}^{r=r_H} \frac{1}{8\pi} t \cdot \boldsymbol{\epsilon} \delta \Lambda.
        \label{dm-odj}
    \end{equation}
    
    From\cite{bardeen_four_1973.Carter.Hawking}, the variation of $\kappa$ is: $$\delta \kappa = -\frac{1}{2}\xi^b \nabla^a \delta g_{ab} + n_a \xi_b \nabla^a \phi^b \delta \Omega_H$$
    The RHS of \eqref{dm-odj} can be evaluated as below:
    \begin{equation}
        \begin{aligned}
            \int_H \bar{\delta} \boldsymbol{Q}_\xi & = \int_H \delta \boldsymbol{Q}_\xi - \int_H \boldsymbol{Q}[\delta\xi] \\
                &   = -\frac{1}{16\pi} \delta \int_H  \left(\nabla^d \xi^b + 2 A_a \xi^a F^{db} \right) \boldsymbol{\epsilon}_{mndb} - \int_H  \left(\nabla^d \delta\Phi_H \phi^b + 2 A_a \delta\Phi_H \phi^a F^{db} \right) \boldsymbol{\epsilon}_{mndb}\\
                &   =  \delta(\frac{\kappa A}{8 \pi} + \Phi_H Q) + \delta \Omega_H J
                \label{bardeltaQ}
        \end{aligned}
    \end{equation}
    The second line uses \eqref{Q_A}, the first term of third line uses the result $\int_H \nabla^a \xi^b \epsilon_{cdab} = -2 \kappa A $ from \cite{wald2010GR}, the second term of the third line comes from defining $\Phi_H := -A_a \xi^a|_H $ and $Q := \frac{1}{8\pi} \int_H F^{ab} \epsilon_{abcd}$, 
    the relation $\epsilon_{abcd} = \hat{\epsilon}_{ab}\wedge \xi_c \wedge n_d$ is used to perform the integrals on horizon where $\hat{\epsilon}_{ab}$ is surface element 2-form on horizon and $n_a$ is defined as $n_a \xi^a =-1$.
    \begin{equation}
        \begin{aligned}
            \int_H  \xi \cdot \boldsymbol{\Theta} & = \int_H \xi^l \frac{1}{16 \pi} \left[g^{ce} g^{ab}\left(\nabla_b \delta g_{ac} - \nabla_c \delta g_{ab}\right) - 4 F^{eb} \delta A_b\right] \epsilon_{mnel}\\
            & =\int_H  \frac{-1}{16\pi} \left(\xi^c \nabla^a \delta g_{ac} - \xi^c \nabla_c(g^{ab} \delta g_{ab}) - 4 \xi_e F^{eb} \delta A_b\right)\hat{\epsilon}_{mn}\\
            & = \int_H \frac{-1}{16 \pi} \left(-2 \delta \kappa + 2 n_a \xi_b \nabla^a \phi^b \delta \Omega_H + 4 F^{eb}n_b\xi_e \xi^b \delta A_b\right)\hat{\epsilon}_{mn}\\
            & = \frac{1}{8\pi} A \delta \kappa + \frac{1}{16\pi} \int_H \epsilon_{mnab} \nabla^a \phi^b \delta \Omega_H - \int_H \frac{1}{4\pi} F^{eb} n_b \xi_e \xi^b \delta A_b
        \end{aligned}
    \end{equation}
    the third term of the third line comes from the fact that $\xi_e F^{ef} \propto \xi$(noticed in \cite{Gao_2001})
    \begin{equation}
        \begin{aligned}
            Q \delta \Phi_H = -\frac{1}{8\pi} \int_H F^{ab}\epsilon_{cdab} A_e \phi^e \delta \Omega_H  -  \frac{1}{4\pi} \int_H\hat{\epsilon}_{cd} F^{ab} \xi_a n_b \delta A_e \xi^e
        \end{aligned}
    \end{equation}
    Then, by applying the definition of Noether charge(especially with Killing vector $\pdv{}{\phi}$), it is directly to notice:
    \begin{equation}
        \begin{aligned}
            \int_H \xi \cdot \boldsymbol{\Theta} - Q \delta \Phi_H = \frac{1}{8 \pi} A \delta \kappa + \delta \Omega_H J.
            \label{xiTheta}
        \end{aligned}
    \end{equation}
    Finally, bring \eqref{bardeltaQ} and \eqref{xiTheta} back to \eqref{dm-odj}, and define: 
    \begin{equation}
        \begin{aligned}
            \delta P & := -\frac{\delta \Lambda}{8 \pi} ,\\
            V &:= \int_{r=0}^{r=r_H}t \cdot \boldsymbol{\epsilon} 
        \end{aligned}
    \end{equation}
    then one gets first law with $P-V$ term:
    \begin{equation}
        \delta M = \frac{\kappa}{8\pi}\delta A +\Omega_H \delta J +\Phi_H \delta Q + V \delta P.
        \label{FirstLaw_A}
    \end{equation}

    And the exact definition of all thermodynamic quantities except for mass are listed below:

    \begin{equation}
    \begin{aligned}
        A & := \int_{S} \sqrt{\sigma}, & \kappa & := \sqrt{-\frac{1}{2} \nabla^{a} \xi^b \nabla_{a} \xi_b }, \\
        J &:= -\int_S \boldsymbol{Q}_\psi, & \Omega &:= - \frac{g_{t \psi}}{g_{\psi \psi}} |_H; \\
        Q &:= \frac{1}{8\pi} \int_H F^{ab} \epsilon_{abcd}, & \Phi_H &:= -A_a \xi^a|_H , \\
        V &:= \int_{r=0}^{r=r_H}t \cdot \boldsymbol{\epsilon}, &  P & := -\frac{ \Lambda}{8 \pi} ,
    \end{aligned}
    \label{def thermo quantities_A}
\end{equation}

\newpage
\section{Using Conformal Completion Method to Check The Results}
    There are also other ways to evaluate thermodynamic quantities. 
    In this section, we apply conformal completion method to double check our result. 
    This method was introduced in \cite{A.Ashtekar_1984.A_AdS}, and was developed in \cite{Ashtekar2000} \cite{Das2000conserved}.

    \subsection{A Brief Review of Conformal Completion Method}
    Technically, conformal completion approach is a way to define conservative quantities which satisfies a conservative equation.
    The conformal completion conservative quantities is defined as below:
    \begin{equation}
        Q[\xi] = -\frac{l}{8 \pi} \int_\mathcal{C} \mathcal{E}_{a}^{\: b} \xi^a dS_b, 
        \label{conformal completionQ}
    \end{equation}
    in which, $\mathcal{C}$ is a co-dimension 2 surface at the conformal boundary, also known as "section"; 
    $\mathcal{E}_{ab}$ means the electrical part of asymptotically Weyl tensor($\mathcal{E}_{ab} = lim_{\rightarrow \mathcal{I}}\Omega^{3-d} l^2C_{ambn} n^m n^n;n^a = \nabla^a \Omega$);
    $\xi$ is the Killing vector(or at least asymptotically Killing vector) which is needed in this definition;
    $dS^b$ is the volume of section $\mathcal{C}$.

    \subsection{Evaluate Conformal Mass and Compare with Mass from Iyer-Wald Formalism}
    To evaluate conformal completion mass, all one needs to do is to evaluate every term presented in \eqref{conformal completionQ},then do the integral.

    First, take conformal factor as $\Omega = \frac{1}{r}$. 
    The conformal metric will come to:
    \begin{equation}
        g_{ab} = \Omega^2 \hat{g}_{ab} = (\frac{1}{r})^2 \hat{g}_{ab}.
    \end{equation}
    Here $\hat{g}_{ab}$ represents the physical metric which is used in above sections.
    Under this conformal metric, we can evaluate 
    $$\mathcal{E}^t_{\ t} = lim_{r \rightarrow \infty} l^2 \Omega^{-1} C^t_{\ atb} n^a n^b= 2 m.$$
    
    Second, one need to evaluate $dS_a$. Due to $g_{\psi \psi}|_{\mathcal{I}} = 0$, volume at section cannot directly evaluate as $\sqrt{g_{\theta \theta} g_{\phi \phi}}$.
    A reasonable way to evaluate is in \cite{Chen:2005zj.LuH}\cite{WuDi:2021vch}. Follow the guidance of them, the approach to get $dS_t$ is basically by taking the limit of $r \rightarrow \infty$ to get the metric od conformal boundary and evaluate the volume element associated with it.
    Then use normal vector to contract with the boundary volume element.
    
    The boundary metric is:
    \begin{equation}
        \gamma_{ab} = lim_{r \rightarrow \infty}g_{ab} = -\frac{1}{l^2}dt_a dt_b + 2 \frac{sin^2 \theta}{l}dt_a d\psi_b +  \frac{1}{sin^2\theta}d\theta_a d\theta_b ,
    \end{equation}
    and the associated volume element is:
    \begin{equation}
        \boldsymbol{\epsilon} = \frac{sin^2 \theta}{L^2} dt_a \wedge d\theta_b \wedge d\psi_c.
    \end{equation}
    Take $dS_t$ to be:
    \begin{equation}
        \left(\pdv{}{t}\right) \cdot \boldsymbol{\epsilon} = \frac{sin^2 \theta}{L^2} d\theta_b \wedge d\psi_c
    \end{equation}
    Finally, it can be directly integrated that the conformal completion mass associated with Killing vector:$\pdv{}{t}^a$ is:
    \begin{equation}
        \mathcal{M}_{C} = \frac{\mu m}{2\pi} ,
    \end{equation}
    which is compatible with early result.

    Conformal mass and angular momentum in any other dimension can be evaluated in same way.

\newpage

\bibliography{refs}

\end{document}